\documentclass[aps, prd, twocolumn, superscriptaddress,showpacs,nofootinbib,notitlepage, bibliography]{revtex4-2}
\usepackage{amsmath}
\usepackage{amssymb} 
\usepackage{graphicx}
\usepackage{natbib}
\usepackage{enumerate}
\usepackage{color}
\usepackage{tensor}
\usepackage{mathrsfs}
\usepackage{hyperref}
\usepackage[normalem]{ulem}
\usepackage[dvipsnames]{xcolor}
\usepackage{accents}
\usepackage{scalerel}
\usepackage{bm}
\usepackage{anysize}
\usepackage{siunitx}
\usepackage{longtable}
\usepackage{dcolumn}

\usepackage{multirow}

\makeatletter
\newcommand*{\leqdef}{\mathrel{\rlap{%
			\raisebox{0.25ex}{$\m@th\cdot$}}%
		\raisebox{-0.25ex}{$\m@th\cdot$}}%
	=}
\newcommand*{\reqdef}{=\mathrel{\rlap{%
			\raisebox{0.25ex}{$\m@th\cdot$}}%
		\raisebox{-0.25ex}{$\m@th\cdot$}}
}

\newcommand{\NSBHfull}{\texttt{IMRPhenomNSBH}}
\newcommand{\BNSfull}{\texttt{IMRPhenomPv2\_NRTidal}}
\newcommand{\Bilby}{\texttt{Bilby}}

\newcommand{\emcee}{\texttt{emcee}}
\newcommand{\getdist}{\texttt{getdist}}

\makeatother

\begin{document}

\title{Prospects of constraining $f(T)$ gravity with the third-generation gravitational-wave detectors}
\author{Ran Chen}
\affiliation{Key Laboratory of Dark Matter and Space Astronomy, Purple Mountain Observatory, Chinese Academy of Sciences, Nanjing 210033, People's Republic of China}
\affiliation{School of Astronomy and Space Science, University of Science and Technology of China, Hefei, Anhui 230026, People's Republic of China}
\author{Yi-Ying Wang}
\affiliation{Key Laboratory of Dark Matter and Space Astronomy, Purple Mountain Observatory, Chinese Academy of Sciences, Nanjing 210033, People's Republic of China}
\affiliation{School of Astronomy and Space Science, University of Science and Technology of China, Hefei, Anhui 230026, People's Republic of China}
\author{Lei Zu}
\affiliation{Key Laboratory of Dark Matter and Space Astronomy, Purple Mountain Observatory, Chinese Academy of Sciences, Nanjing 210033, People's Republic of China}
\author{Yi-Zhong Fan}
\email{yzfan@pmo.ac.cn}
\affiliation{Key Laboratory of Dark Matter and Space Astronomy, Purple Mountain Observatory, Chinese Academy of Sciences, Nanjing 210033, People's Republic of China}
\affiliation{School of Astronomy and Space Science, University of Science and Technology of China, Hefei, Anhui 230026, People's Republic of China}

\begin{abstract}
Mergers of binary compact objects, accompanied with electromagnetic (EM) counterparts, offer excellent opportunities to explore varied cosmological models, since gravitational waves (GWs) and EM counterparts always carry the information of luminosity distance and redshift, respectively.
$f(T)$ gravity, which alters the background evolution and provides a friction term in the propagation of GWs, can be tested by comparing the modified GW luminosity distance with the EM luminosity distance. Considering the third-generation gravitational-wave detectors, Einstein Telescope and two cosmic explorers, we simulate a series of GW events of binary neutron stars and neutron-star–black-hole binaries with EM counterparts. These simulations can be used to constrain $f(T)$ gravity [especially the power-law model $f(T)=T+\alpha(-T)^\beta$ in this work] and other cosmological parameters, such as $\beta$ and the Hubble constant. In addition, combining simulations with current observations of type Ia supernovae and baryon acoustic oscillations, we obtain tighter limitations for $f(T)$ gravity. We find that the estimated precision significantly improved when all three datasets are combined ($\Delta \beta \sim 0.03$), compared to analyzing the current observations alone ($\Delta \beta \sim 0.3$). Simultaneously, the uncertainty of the Hubble constant can be reduced to approximately $1\%$. 

\end{abstract}

\maketitle

\section{Introduction}
\label{sec:intro}
General relativity (GR) has been proven by many experiments, including the recent direct detection of gravitational waves (GWs) from the mergers of two compact objects \cite{LIGOScientific:2016aoc, LIGOScientific:2017vwq}.
However, the presence of singularities at the centers of black holes and the breakdown of the equivalence principle at the singularities indicate that GR is not a universal theory of spacetime \cite{Shankaranarayanan:2022wbx}, and the theoretical and observational challenges faced by the $\Lambda$ cold dark matter ($\Lambda$CDM) model might suggest a new theory of gravity~\cite{Clifton:2011jh, Dolgov:2003px}.
In addition to dark energy, many modified gravity models have been developed to account for the accelerating expansion of the Universe. Among these models, one prominent approach involves incorporating the curvature through $f(R)$ gravity~\cite{Sotiriou:2008rp,DeFelice:2010aj}. 
As an extension of the teleparallel equivalent of general relativity (TEGR)~\cite{Maluf:2013gaa}, $f(T)$ gravity represented the most general modified form of TEGR and garnered considerable attention in previous studies~\cite{Cai:2015emx,Bahamonde:2021gfp,Krssak:2018ywd}.
Diverging from $f(R)$ gravity, characterized by curvature, $f(T)$ gravity utilizes torsion as a geometric object to describe gravity. In addition, $f(T)$ gravity is a second-order theory due to the fact that the torsion scalar $T$ depends only on the first derivatives of the  vierbein~\cite{Bahamonde:2015zma}.
Some studies have constrained $f(T)$ gravity by observations, such as the investigation with GW170817/$\gamma$-ray burst (GRB) 170817A~\cite{Cai:2018rzd}, the corresponding limitations based on cosmic microwave background observations, including the effects of primordial gravitational waves on the BB spectrum~\cite{Nunes:2018evm}, and the impact of the linear scalar perturbations evolution on the TT anisotropy power spectrum~\cite{Kumar:2022nvf}, etc. (see Refs.~\cite{Nunes:2019bjq,Zhang:2021kqn,Tzerefos:2023mpe,Briffa:2023ern} for comprehensive reviews).

Within the framework of modified gravity and in the absence of anisotropic stress, the tensor perturbations bring the modifications of the GW propagation equation~\cite{Nishizawa:2017nef}.
In general, these modifications can alter the dispersion relationship, resulting in deviations between the speed of GW propagation and the speed of light. 
Recent studies of the GW170817/GRB 170817A event~\cite{LIGOScientific:2017zic,Wang:2017rpx}
provided a tight constraint $\lvert c_{\text{gw}}/c - 1 \rvert < 10^{-15}$ at low redshifts.
However in $f(T)$ gravity, the sole modification in the equation of GW propagation is the friction term, which does not change the speed of GW propagation.
Moreover, extensive investigations \cite{Belgacem:2018lbp,Ferreira:2022jcd,Matos:2021qne,Finke:2021aom,LISACosmologyWorkingGroup:2019mwx,Mastrogiovanni:2020mvm,2023arXiv230409025Z} have been devoted to exploring the apparent difference between the GW luminosity distance ($d_{\rm L}^{gw}$) and electromagnetic (EM) luminosity distance ($d_{L}^{em}$), which result from the friction term in the GW propagation function of modified gravity.
Meanwhile, the effective field theory of $f(T)$ gravity within the cosmological framework presents an opportunity to modify the evolution of Hubble parameter $H(z)$, offering a model of late-time effective dark energy and releasing the $H_0$ tension~\cite{Ren:2022aeo,Wang:2020zfv}. Besides, GW standard sirens present an independent and distinctive approach to investigating the $H_0$.

In this paper, we present an end-to-end analysis of the simulated binary neutron stars (BNSs) and neutron-star–black-hole (NSBH) binary samples with the third-generation gravitational-wave observatories, Einstein Telescope (ET) and Cosmic Explorer (CE). These prospective detectors are expected to exhibit significant improvements in sensitivity compared to current ground-based gravitational-wave detectors during the O5 observing period ~\cite{Maggiore:2019uih,Sathyaprakash:2009xt,2021arXiv210909882E}, thereby providing a valuable opportunity to constrain $f(T)$ gravity at high redshifts. 
Because of the degeneracy between the inclination angle and the luminosity distance in GW observations, we construct different population models for BNS and NSBH events, including the presence of short $\gamma$-ray bursts (SGRBs). Such EM counterpart improves the precisions of estimated $d_{\rm L}^{gw}$ \cite{2023ApJ...943...13W}, ultimately enhancing our ability to constrain $f(T)$ gravity from GW events. 
To establish the baselines of simulated GWs, we begin by constraining various cosmological parameters using current observations, specifically type Ia supernovae (SN Ia) and baryon acoustic oscillations (BAOs). This initial analysis allows us to obtain reliable cosmological constraints. Then, we discuss the capability of constraining $f(T)$ gravity with the simulated GW catalog. By combining the information from SN Ia, BAOs, and the simulated GW events, we find that the precision of constraining $f(T)$ gravity is significantly improved. This enhanced precision offers an opportunity to explore the possible deviations from GR within the framework of $f(T)$ gravity.
It is worth noting that, compared with previous works~\cite{Nunes:2019bjq,Zhang:2021kqn}, we have conducted a comprehensive analysis of simulated GW events, including the specific parameters of the binaries, and the conditions for generating the electromagnetic counterpart, etc. 
In addition, our simulations select a special class of GW events, which have small inclination angles and are accompanied by on axis GRBs. In this case, the inclination angle can be limited to $\Delta \iota \le 0.1$ rad directly. Therefore, the estimations of GW luminosity distance $\Delta d_{\rm L}$ will be more accurate, because the degeneracy between the inclination angle and the luminosity distance in the GW analysis can be broken.
 
The paper is organized as follows: In Sec.~\ref{sec:f(T)}, we provide a brief review of $f(T)$ gravity and discuss the modified propagation of GW within this framework. We focus on the power-law model, which is one of the most commonly used models for $f(T)$.
In Sec.~\ref{sec:Observational baselines}, we describe the baselines of GW simulations, which are based on realistic observations of SN Ia and BAOs. 
In Sec.~\ref{sec:Modeling the gravitational signal}, we generate a simulated catalog of BNS and NSBH events as standard sirens, with detailed physical motivation and special selection criteria. Additionally, we analytically calculate the prospective accuracy of the luminosity distance estimation from GWs using the combination of ET+2CE. 
In Sec.~\ref{sec:result and discussion}, we present the main results and engage in in-depth discussions regarding the constraints of $f(T)$ gravity. 
Finally, we summarize the key findings of the study and present our conclusions in Sec.~\ref{sec:Conclusion}.

\section{$f(T)$ gravity}\label{sec:f(T)}
In this section, we provide a brief overview of $f(T)$ gravity, along with an analytical treatment of the evolution of the Hubble parameter and modified GW propagation within the framework of the power-law model.

\subsection{Dynamical evolution}
The dynamical variable of the teleparallel gravity and its $f(T)$ extension is the vierbein field $\mathbf{e}_A=e_A^\mu \partial_\mu$ and the action of $f(T)$ gravity takes the following form~\cite{Cai:2015emx}:
\begin{equation}
	\begin{aligned}
        \mathcal{S}=\int \left[\frac{1}{16 \pi G}f(T)+L_{m}\right]|e|d^4 x,
	\end{aligned}
\end{equation}
where $L_{m}$ is the Lagrangian of matter and $|e|=\operatorname{det}\left(e_\mu^A\right)=\sqrt{-g}$.
We express $f(T)$ in the form of $f(T)=T+F(T)$ and denote $\Xi_{T}= \partial \Xi / \partial T, \Xi_{T T}=\partial^2 \Xi / \partial T^2$, where $\Xi$ represents $f$ and $F$, respectively.

Assuming the cosmology is dominated by the flat Friedmann-Robertson-Walker (FRW) metric
\begin{equation}
	\begin{aligned}
        d s^2=d t^2-a^2(t) \delta_{i j} d x^i d x^j,
	\end{aligned}
\end{equation}
which can be obtained by taking the vierbein as $e_\mu^A=\operatorname{diag}(1, a, a, a)$ with the scale factor $a$.

\begin{figure}[t]
\centering
\includegraphics[width=0.45\textwidth]{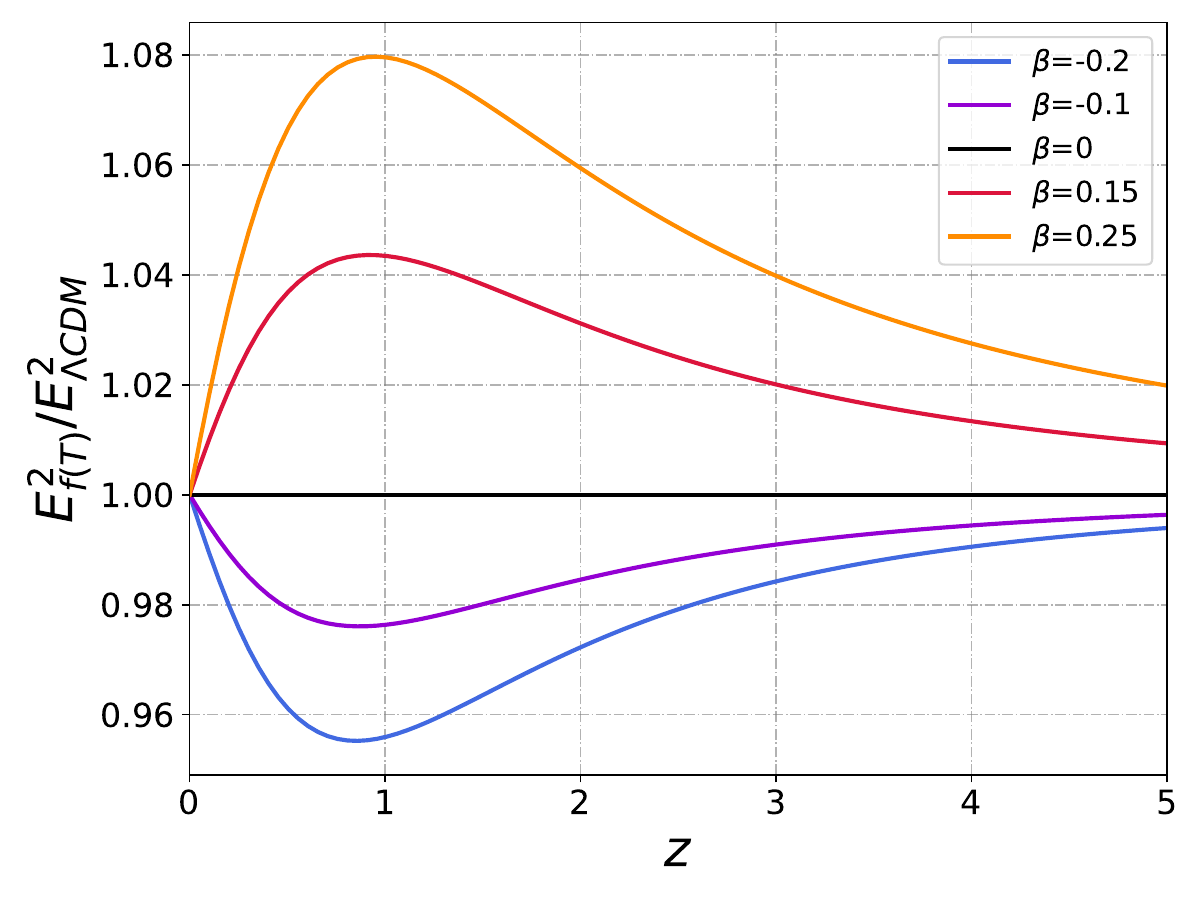}
\caption{The ratio of the normalized Hubble parameter $E^{2}_{f(T)}/E^{2}_{\Lambda \rm CDM}$ at different parameters of $f(T)$ gravity as a function of the redshift $z$.}
\label{fig:efunc}
\end{figure}

Variation of the action with respect to the tetrad $e_\nu^A$ leads to the Friedmann equations~\cite{Cai:2015emx}:
\begin{equation}
	\begin{aligned}
        3 H^2 & =8 \pi G \rho_m-\frac{F(T)}{2}+T F_T \\
        \dot{H} & =-\frac{4 \pi G\left(\rho_m+p_m\right)}{1+F_T+2 T F_{T T}},
	\end{aligned}
\end{equation}
where $H \equiv \dot{a} / a$ is the Hubble parameter, $p_m$ is the pressure density and $\rho_m$ is the energy density for the matter fluid.

Once the vierbein is obtained, the relation for the torsion scalar can be written as
\begin{equation}
	\begin{aligned}
        T & =-6 H^2 \\
        T & \equiv S_\rho{ }^{\mu \nu} T^\rho{ }_{\mu \nu},
	\end{aligned}
\end{equation}
where $T^\rho{ }_{\mu \nu}=e_A^\rho\left(\partial_\mu e_\nu^A-\partial_\nu e_\mu^A\right)$ and $S_\rho{ }^{\mu \nu}=\frac{1}{2}\left(\mathcal{K}^{\mu \nu}{ }_\rho+\delta_\rho^\mu T^{\alpha \nu}{}_\alpha-\delta_\rho^\nu T^{\alpha \mu}{}_\alpha\right)$ represent the torsion tensor and the superpotential, respectively. 
The contortion tensor, denoted as $\mathcal{K}^{\mu \nu}{}_\rho=-\frac{1}{2}\left(T^{\mu \nu}{}_\rho-T^{\nu \mu}{}_\rho-T_\rho{}^{\mu \nu}\right)$, represents the divergence of the Levi-Civita connection in Riemannian spacetime and the Weitzenböck connection in Weitzenböck spacetime~\cite{Maluf:2013gaa,Cai:2015emx}.

\subsection{Power-law model}
As an alternative explanation for the observed accelerated expansion of the Universe at late times, $f(T)$ gravity incorporates the energy density and pressure as an ``effective dark energy"~\cite{Cai:2015emx} component arising from the gravitational term.

The power-law model~\cite{Bengochea:2008gz} of $f(T)$ gravity reads 
\begin{equation}
	\begin{aligned}
		&f(T) =T+\alpha (-T)^\beta, \\
        &\alpha =\left(6 H_0^2\right)^{1-\beta} \frac{(1-\Omega_{m}^{(0)}-\Omega_{r}^{(0)})}{2 \beta-1}, 
	\end{aligned}
\end{equation}
where $\Omega_m^{(0)}=\frac{8 \pi G }{3 H_0^2}\rho_m^{(0)}, \Omega_r^{(0)}=\frac{8 \pi G }{3 H_0^2}\rho_r^{(0)}$ and they represent the matter and radiation density parameter of the current Universe, respectively.

The evolution equation of $ H(z)$, with the definition of the normalized Hubble parameter denoted by $E^2(z)$, is expressed as~\cite{Li:2018ixg}
\begin{equation}\label{eq:efun}
	E^{2} \equiv \frac{H^2}{H_0^2}=\Omega_{m}^{(0)}(1+z)^3+\left(1-\Omega_{m}^{(0)}\right) E^{2\beta}.
\end{equation}
Here, we ignored the radiation density parameter of the current Universe (i.e., $\Omega_{r}^{(0)}=0$). Figure.~\ref{fig:efunc} showed the evolution of Eq.~(\ref{eq:efun}) compared with the $\Lambda \mathrm{CDM}$ model ($\beta = 0$). The second Friedmann equation can be written as
\begin{equation}\label{eq:H_dot_H2}
\frac{\dot{H}}{H^2}=-\frac{3}{2}\left[\frac{1- (1-\Omega_{m}^{(0)})E^{2\beta-2}}{1-\beta(1-\Omega_{m}^{(0)})E^{2\beta-2}}\right],
\end{equation}
where $\dot{H}$ denotes a derivative with respect to the cosmic time.
Particularly, for the $\Lambda \mathrm{CDM}$ model ($\beta=0$), Eq.~(\ref{eq:H_dot_H2}) degenerates to 
\begin{equation}
\left(\frac{\dot{H}}{H^{2}}\right)_{\Lambda \rm CDM} = -\frac{3}{2}\left[1-\frac{1}{E^{2}}\left(1-\Omega_{m}^{(0)}\right)\right].
\end{equation}

\subsection{GW propagation in $f(T)$ gravity}

The perturbation part of the tensor sectors of $f(T)$ gravity around the FRW cosmological background leads to the equation of motion for the GW, which is consistent with the result from Arnowitt-Deser-Misner decomposition of the vierbein field~\cite{Chen:2010va,Cai:2018rzd,Tzerefos:2023mpe}
\begin{equation}\label{ftpro}
h_{\boldsymbol{\lambda}}^{\prime \prime}+2 \mathcal{H}\left(1-\beta_T\right) h_{\boldsymbol{\lambda}}^{\prime}+k^2 h_{\boldsymbol{\lambda}}=0,
\end{equation}
where the prime denotes derivative with respect to the conformal time and $\lambda$ represents the two polarization states of the tensor modes~\cite{Farrugia:2018gyz}. The dimensionless parameter $\beta_T$ is defined as
\begin{equation}\label{eq:beta}
	\begin{aligned}
		\beta_T=-\frac{f_{T}^{\prime} }{2 \mathcal{H} f_T}=- \frac{\dot{f_{T}}}{2 H f_T}=\frac{6f_{TT}\dot{H}}{f_{T}}.
	\end{aligned}
\end{equation}
By resubstituting the scale factor $\frac{\tilde{a}^{\prime}}{\tilde{a}}=\mathcal{H}[1-\beta_T]$ and performing some straightforward derivations, the redefined GW luminosity distance is related to the electromagnetic luminosity distance ~\cite{Belgacem:2017ihm}
\begin{equation}\label{dl_GW}
	\begin{aligned}
		d_L^{gw}(z)=d_L^{\mathrm{em}}(z) \exp \left[-\int_0^z d z^{\prime} \frac{\beta_T\left(z^{\prime}\right)}{1+z^{\prime}}\right].
	\end{aligned}
\end{equation}
In the power-law model of $f(T)$ gravity, we introduce a simplifying factor $\kappa = \frac{(4-2\beta^{-1})}{3(\Omega_{m}^{(0)}-1)}$ and denote $\beta_T$ with $z$ as its variable. By substituting Eqs.~(\ref{eq:efun}) and~(\ref{eq:H_dot_H2}) into Eq.~(\ref{eq:beta}), we obtain the expression of $\beta_T(z)$ as follows:
\begin{equation}\label{delta}
	\beta_T(z) = \frac{(\beta -1)\left[1 + (\Omega_{m}^{(0)}-1)E^{2\beta-2} \right]}{\kappa(\beta+E^{2-2\beta})+\left[1+\beta(\Omega_{m}^{(0)}-1)E^{2\beta-2}\right]}.
\end{equation}

\begin{figure}[t]
\centering
\includegraphics[width=0.45\textwidth]{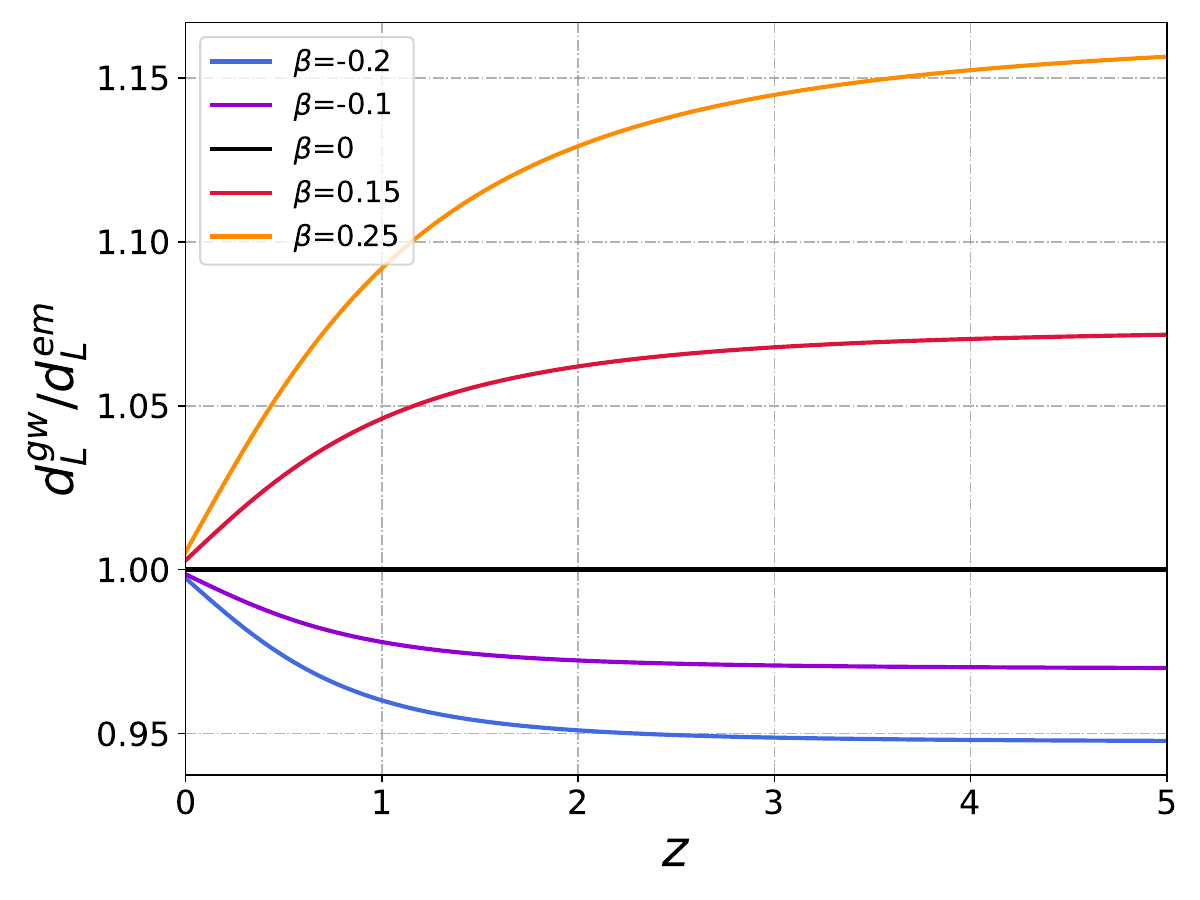}
\caption{The evolution of the ratio $d_L^{gw} / d_L^{em}$ with increasing redshift. Different colored lines illustrate the deviation from GR with different $f(T)$ gravity parameters.}
\label{fig:dl_gw_em}
\end{figure}

To quantify the deviation of $f(T)$ gravity from GR, we plot the ratio $d_{L}^{gw}/d_{L}^{em}$ as a function of redshift $z$, considering different parameters of $f(T)$ gravity (see Fig.~\ref{fig:dl_gw_em}). It is evident that this deviation becomes more pronounced at higher redshifts.

\section{Current observations and GW-simulation baselines}\label{sec:Observational baselines}
$f(T)$ gravity provides a plausible explanation for cosmic evolution in various aspects by introducing effective dark energy during the late stages of the Universe. It also changes structure formation in the early Universe. This framework offers a promising avenue to alleviate the Hubble tension~\cite{Ren:2022aeo,Wang:2020zfv}. In this section, we introduce the observations of SN Ia and BAOs, which can be used to constrain the $f(T)$ gravity.

\subsection{Type Ia supernovae}
Type Ia supernovae have been widely recognized as standard candles for measuring cosmic acceleration in the local Universe~\cite{Branch:1992rv,Wright:2017rsu}. In our analysis, we utilize the Pantheon catalog, which comprises 1048 supernovae spanning the redshift range $0.01 < z < 2.3$~\cite{Pan-STARRS1:2017jku}.
At special redshift $z_i$, the apparent magnitude of an SN Ia is 
\begin{equation}
    m=5 \log _{10}\left[D_L\left(z_i, \Theta\right)\right]+25 + M, 
\end{equation}
where $D_L\left(z_i, \Theta\right)$ is the luminosity distance in $f(T)$ gravity, which can be calculated using  
\begin{equation}
D_L\left(z_i, \Theta\right)=c\left(1+z_i\right) \int_0^{z_i} \frac{d z^{\prime}}{H}\left(z^{\prime}, \Theta\right),
\end{equation}
where {\bf $H$} corresponds to the Hubble parameter in $f(T)$ gravity as defined in Eq.~(\ref{eq:efun}). 
Additionally, the apparent magnitude of each SN Ia needs to be calibrated using an arbitrary fiducial absolute magnitude $M$. In our Markov chain Monte Carlo analyses, we treat $M$ as a prior parameter and then marginalize it. 

The likelihood function for SN Ia observations can be represented as 

\begin{equation}
	\begin{aligned}
    &\chi_{\mathrm{sne}}^2\left(\Theta, M\right)= \\
    &\sum_{i j}\left[m_i-m\left(z_i\right)\right] S_{\mathrm{sne}, i j}^{-1}\left[m_j-m\left(z_j\right)\right],
    \end{aligned}
\end{equation}
where $m(z_i)$ is the apparent magnitude at $z_i$, and $S_{\text{sne}}$ is the covariance matrix that accounts for both the statistical and systematic uncertainties associated with the Pantheon catalog.

\begin{table*}[t] 
	\centering
	\caption{Injection parameter distributions of the BNS and NSBH populations}
	\label{tab:Injection parameters}
	\setlength{\extrarowheight}{8pt}
	\begin{tabular}{l|cc|cc|cc|cc|cc}\hline 
		\hline \multirow{2}{*}{ Parameter } & \multicolumn{2}{c|}{ BNS } & \multicolumn{2}{c}{ NSBH } \\ \cline{2-5}
		 & \multicolumn{2}{c|}{ Neutron Stars } & \multicolumn{1}{c}{ Neutron Stars } & \multicolumn{1}{c}{ Black Hole } \\

		\hline Mass $m$ & \multicolumn{2}{c|}{$\mathcal{N}\left(1.33,0.11^2\right)$}  & \multicolumn{1}{c}{$ \mathcal{N}\left(1.6,0.11^2\right)  \mathrm{M}_{\odot}$} & \multicolumn{1}{c}{$[5,30]  \mathrm{M}_{\odot}$} \\
		
		\hline Mass model & \multicolumn{2}{c|}{ Gaussian } & \multicolumn{1}{c}{Gaussian} & \multicolumn{1}{c}{fiducial model~\cite{Zhu:2020ffa}} \\
		
		\hline $\operatorname{Spin} \chi$ & \multicolumn{2}{c|}{ $[0,0.05]$~\cite{LIGOScientific:2018hze}}  & \multicolumn{1}{c}{$[0,0.05]$} & \multicolumn{1}{c}{$ \mathcal{N}\left(0.8,0.15^2\right) $} \\
		
		\hline Spin model & \multicolumn{2}{c|}{ Aligne Uniform } & \multicolumn{1}{c}{Aligne Uniform} & \multicolumn{1}{c}{Gaussian} \\
		
		\hline Waveform & \multicolumn{2}{c|}{ \BNSfull~\cite{Dietrich:2018uni}} & \multicolumn{2}{c}{\NSBHfull~\cite{Thompson:2020nei}}  \\
		
		\hline Tidal deformability $\Lambda_{NS,BH}$ & \multicolumn{4}{c}{ $\Lambda_{NS}$:Computed from AP4 EOS, $\Lambda_{BH}$:0} \\
		
		\hline Redshift $z$ & \multicolumn{4}{c}{ Madau-Dickinson $+P\left(t_d\right) \propto 1 / t_d, t_{d, \min }=20 \mathrm{Myr}$~\cite{Iacovelli:2022bbs}} \\
		
		\hline Luminosity distance $D_L$ & \multicolumn{4}{c}{ Computed from $z$ assuming flat $\Lambda \rm CDM$ $\left\{H_{0}=68.6, \Omega_m^{(0)}=0.30 \right\}$} \\
		
		\hline Inclination angle $\theta_{\rm j}$ & \multicolumn{4}{c}{ Sine in [0,0.1]} \\
		
		\hline Right ascension $\alpha$ & \multicolumn{4}{c}{ Uniform in [0,2$\pi$] } \\
		
		\hline Declination $\delta$ & \multicolumn{4}{c}{ Cosine } \\

		\hline Polarization of GW $\psi$ & \multicolumn{4}{c}{ Uniform in [0,$\pi$] } \\
		
		\hline Coalescence phase $\phi_c$ & \multicolumn{4}{c}{Uniform in [0,2$\pi$]} \\
		
		\hline Coalescence time $t_c$ & \multicolumn{4}{c}{0} \\
		\hline
		\hline
	\end{tabular}
\end{table*}

\subsection{Baryon acoustic oscillations}

The measurements of BAOs have achieved an accuracy of $3\%$ across 14 narrow redshift shells. These measurements are based on a detailed analysis of Galaxy data from BOSS data release 12 and eBOSS data release 16 in the range of redshift $0.32 < z < 0.66$~\cite{BOSS:2012dmf, Dawson:2015wdb, Sanchez:2010zg, Carvalho:2017tuu, Menote:2021jaq}.
In our analysis, we primarily focus on the peak of the angular correlation function. The angular diameter distance is
\begin{equation}
D_A\left(z_i, \Theta\right)=\frac{c}{\left(1+z_i\right)} \int_0^{z_i} \frac{d z^{\prime}}{ H}\left(z^{\prime}, \Theta\right).
\end{equation}
Then, the BAO peak is at 
\begin{equation}
\theta_{\mathrm{BAO}}(z_i
)=\frac{r_d}{(1+z_i) D_A\left(z_i, \Theta\right)},
\end{equation}
where $r_d$ is the sound horizon at the drag epoch.
Further, the likelihood function for BAO measurement is
\begin{equation}
\begin{aligned}
\chi_{\mathrm{BAO}}^2  \left(\Theta, r_d\right)&= \sum_{i j}\left[\theta_{\mathrm{BAO}}\left(z_i\right)-\theta_{\mathrm{BAO}, i}\right]\\
&\quad\times\Sigma_{\mathrm{BAO}, i j}^{-1}\left[\theta_{\mathrm{BAO}}\left(z_j\right)-\theta_{\mathrm{BAO}, j}\right],
\end{aligned}
\end{equation}
where $\Sigma_{\mathrm{BAO}, i j}$ is the covariance matrix that quantifies the statistical and systematic uncertainties.

\section{Modeling the gravitational signal}
\label{sec:Modeling the gravitational signal}

The merger rates of BNSs and NSBH can be estimated from the GW transient catalog 1–3. The estimated BNS merger rate  is $10–1700$ $\mathrm{Gpc}^{-3} \mathrm{yr}^{-1}$, while the NSBH merger rate is $7.8–140$ $\mathrm{Gpc}^{-3} \mathrm{yr}^{-1}$. 
Previous studies by Jin et al.\cite{2018ApJ...857..128J} found a higher BNS merger rate $452–2541$ $\mathrm{Gpc}^{-3} \mathrm{yr}^{-1}$, using SGRB data. In our work, we apply selection criteria based on the inclination angle, which results in approximately $\int^{0.1}_{0} \sin \iota {\rm d} \iota \sim \iota^2/2 \sim 0.5\%$ events being selected from the total angular distribution. Considering the mean value of the merger rates (855 $\mathrm{Gpc}^{-3} \mathrm{yr}^{-1}$ for BNSs and 74 $\mathrm{Gpc}^{-3} \mathrm{yr}^{-1}$ for NSBH), we expect there will be tens of thousands of BNS and NSBH events available over a one year period, during the epoch with ET+2CE. Specifically, approximately 670 BNS and 50 NSBH events will have inclination angles $\iota \le 0.1 \, \rm rad$. 
Taking into account the proportion of black holes with high spin, which is approximately 24\% \cite{Tang:2022qim}, we estimate that there will be 1000 BNS (100 NSBH) events that can be observed within a two year observation period.
Because of the huge uncertainties associated with merger rates, we will subsequently engage in separate discussions concerning a set of 100 events and another set of 1000 events for BNSs and NSBH, respectively. All of the injection parameters of BNS and NSBH simulations are summarized in \autoref{tab:Injection parameters}, and we formulate them in the following subsections, respectively.

\begin{figure*}[t]
	\centering
	\includegraphics[scale=0.45]{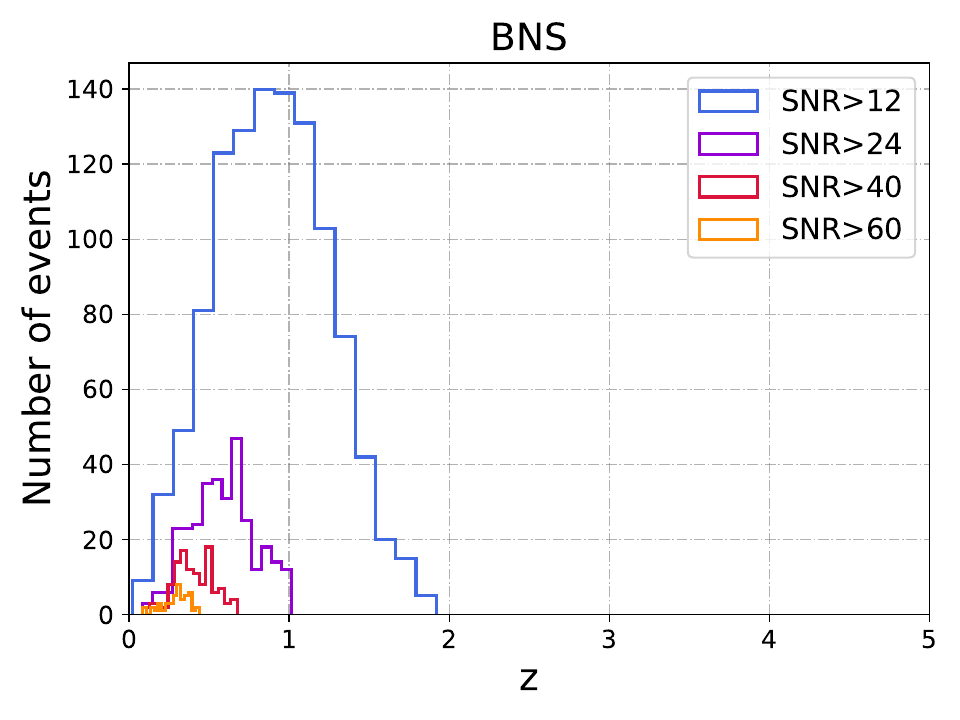}
	\hspace{10pt}
	\includegraphics[scale=0.45]{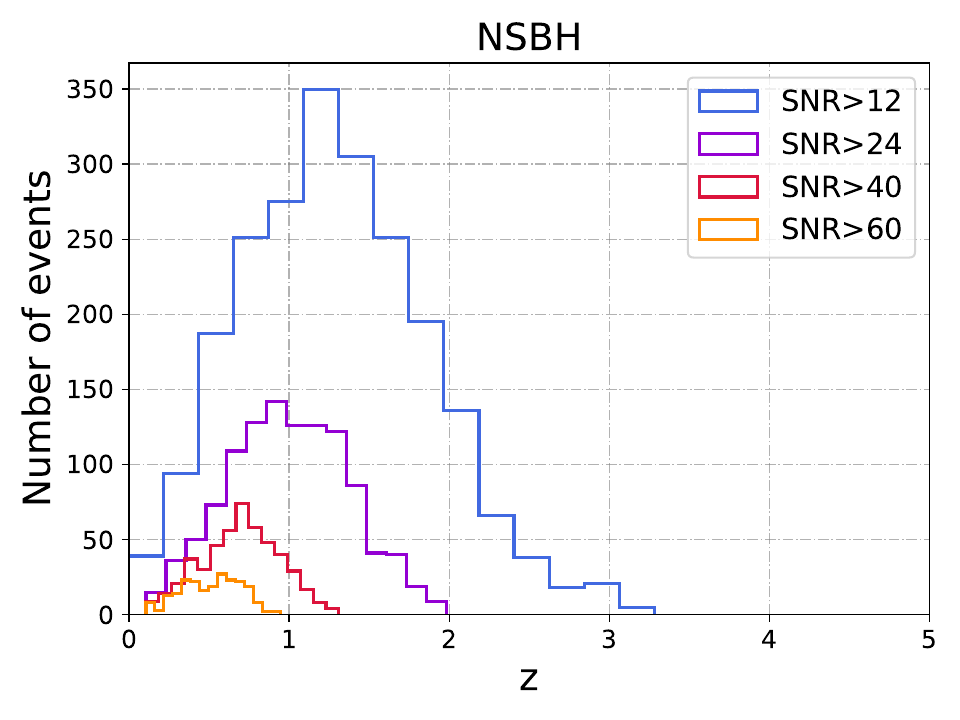}
	\caption{The redshift distribution of the simulated 1000 BNS (left) and 1000 NSBH (right) events with ET+2CE, selected on the basis of different thresholds for the SNR.}
	\label{fig:SNR_distribution}
\end{figure*}

\subsection{BNSs}
Since BNSs are always accompanied with EM counterparts, we generate a population of simulated events in physically motivated parameter spaces. From this simulated population, we select a subset of events with small inclination angles to obtain well-estimated $d_{\rm L}^{gw}$.

(i) Neutron star (NS) mass $M_{\mathrm{NS}}$: The current understanding suggests that neutron stars in BNS systems have masses constrained within a narrow range, peaking at $1.33\mathrm{M}_{\odot}$~\cite{Ozel:2012ax,Kiziltan:2013oja,Shao:2020bzt}.
Therefore, we assume that the distribution of NS mass in BNS systems follows an approximately Gaussian distribution, $M_{\mathrm{NS}} / \mathrm{M}_{\odot} \sim \mathcal{N}\left(1.33,0.11^2\right)$. 

(ii) Equation of state (EOS) and tidal deformability $\Lambda_{NS}$: The AP4 EOS~\cite{Akmal:1998cf} is widely used in nuclear physics to describe the structure and properties of neutron stars. It incorporates various parameters and equations to account for the interactions between nucleons and other particles in dense matter. Relativistic effects and nuclear interactions are considered at high densities to accurately model the behavior of matter.

We employ the AP4 EOS to calculate the tidal deformabilities of neutron stars~\cite{Foucart:2018rjc}, given by
\begin{equation}
    \Lambda=(2 / 3) k_2 C_{\mathrm{NS}}^{-5},
\end{equation}
where $k_2$ is the NS Love number depending on the NS mass, and $C_{\mathrm{NS}}=G M_{\mathrm{NS}} /\left(R_{\mathrm{NS}} c^2\right)$ is the NS compaction determined by the EOS.

(iii) Injected redshift distribution: The redshift distribution of the BNS (NSBH) population follows Madau-Dickinson profile~\cite{Madau:2014bja,Madau:2016jbv}, convolved with a time delay distribution $P\left(t_d\right) \propto 1 / t_d$. Assuming a minimum time delay of 20 Myr, Iacovelli $et$ $al$. \cite{Iacovelli:2022bbs} proposed an analytic equation [Eq.~(A2) therein] to describe the distribution of redshift for these compact objects mergers. Therefore, we utilize it as the redshift distribution for our simulated GW events.

(iv) Inclination angle~$\theta_{\rm j}$: The large uncertainty of $d_{\rm L}^{gw}$ arises from the degeneracy between $d_{\rm L}$ and $\iota$ in the GW events, which significantly impacts the constraints of cosmological models. Fortunately, the viewing angle inferred from SGRBs is widely anticipated to be the same as the inclination angle of the BNS mergers. Therefore, we consider a specific case (on axis) in which the line of sight is within the energetic core of the ejecta (i.e., $\theta_{\rm v} \le \theta_{\rm j}$, $\theta_{\rm j}$ represents the evaluated jet opening angle and we assume $\theta_{\rm j} \sim 0.1$ $\rm{rad}$). The uncertainty of the $\theta_{\rm v}$ (equally, the $\iota$) can be estimated to be within $\sim$ 0.1 rad. Another motivation for considering the on axis case is that the afterglow emissions of off axis SGRBs are highly suppressed until the blast waves driven by the ejecta have decelerated to a bulk Lorentz factor of $\Gamma = 1/\theta_{\mathrm{v}}$ and reached the peak flux at that time. Thus, only the on axis SGRBs have high detectability at high redshifts.

\subsection{NSBH}
Unlike BNS mergers, only a small proportion of NSBH mergers are expected to be accompanied with an EM counterpart. This phenomenon primarily occurs in cases where the stellar-mass black hole has a high spin during the merger event, allowing for the presence of residual baryon mass $M_{\mathrm{rem}}>0$ outside the black hole (BH) after the merger.

(i) BH mass $M_{BH}$ and spin $\chi_{BH}$: Following Refs.~\cite{Mapelli:2018wys,Zhu:2020ffa}, we adopt a formula with two broken exponential decays and an exponential rise to describe the BH mass distribution, i.e., 
$$
\begin{aligned}
f\left(M_{\mathrm{BH}}\right)& \propto  \left(\frac{1}{a_1 \exp \left(-b_1 M_{\mathrm{BH}}\right)+a_2 \exp \left(-b_2 M_{\mathrm{BH}}\right)}\right. \\
& \left.+\frac{1}{a_3 \exp \left(b_3 M_{\mathrm{BH}}\right)}\right)^{-1}\text { for } M_{\mathrm{BH}}>M_{\rm min},
\end{aligned}
$$
where $M_{\rm min}=5\mathrm{M}_{\odot}$ is the minimum BH mass based on the rapid core-collapse supernova model proposed by Ref.~\cite{Fryer_2012}.
The best-fit values for the other parameters are fixed $a_1=1.04 \times 10^{11}$, $b_1=2.1489$, $a_2=799.1$, $b_2=0.2904$, $a_3=0.002845$, and $b_3=1.686$ \cite{Zhu:2020ffa}. 

(ii) NS mass $M_{NS}$ and spin $\chi_{NS}$: As shown in population synthesis studies \cite{Mapelli:2018wys}, the masses of NSs in NSBH systems may be larger than those in BNS systems. The median value of NS masses in NSBH systems is around $1.6\mathrm{M}_{\odot}$. Therefore, we assume a normal distribution $M_{\mathrm{NS}} / \mathrm{M}_{\odot} \sim \mathcal{N}\left(1.6,0.11^2\right)$ with the AP4 NS equation of state (EoS). Regarding the NS spin, we consider a low-spin case since the largest spin of a NS measured in a binary system is 0.02 \cite{Vitale:2015tea,Lorimer:2008se}, and we assume $\chi_{\mathrm{NS}} \sim \text{uniform}\left(0,0.05\right)$.

(iii) Remnant baryon mass: In order to determine whether an EM counterpart is present in a NSBH merger, it is crucial to estimate the remnant baryon mass $\left(M_{\mathrm{rem}}\right)$ outside the black hole after the merger. A positive value of $M_{\mathrm{rem}}$ indicates the formation of a disk that can generate EM counterparts, such as kilonovae, GRBs, and afterglows \cite{Fragione:2021cvv}. On the other hand, if $M_{\mathrm{rem}}\leqslant 0$, it suggests that the NS plunges directly into the black hole without leaving any observable signatures except for the GW.

In the simulations, we focus on the high-spin case for the black hole and assume a normal distribution for the dimensionless spin parameter $\chi_{\mathrm{BH}}$, given by $\chi_{\mathrm{BH}} \sim \mathcal{N}\left(0.8,0.15^2\right)$. This choice is consistent with observations in black hole x-ray binary systems \cite{Qin:2018sxk, McClintock:2011zq}. The remnant disk mass is estimated using the method described in Ref.~\cite{Foucart:2018rjc},
\begin{equation}
    \begin{aligned}
        \hat{M}_{\mathrm{rem}}=\left[\max \left(\alpha \frac{1-2 C_{\mathrm{NS}}}{\eta^{1 / 3}}-\beta \hat{R}_{\mathrm{ISCO}} \frac{C_{\mathrm{NS}}}{\eta}+\gamma, 0\right)\right]^\delta,
    \end{aligned}
\end{equation}
where $(\alpha, \beta, \gamma, \delta)=(0.406,0.139,0.255,1.761)$ are obtained from numerical relativity simulations ~\cite{Foucart:2012nc,Foucart:2012vn}, $\eta=Q /(1+Q)^2$ is the symmetric mass, with $Q=M_{\mathrm{BH}} / M_{\mathrm{NS}}$, and $\hat{R}_{\rm ISCO} = R_{\rm ISCO}/M_{\rm BH}$ is the normalized innermost stable circular orbit (ISCO) radius.

\subsection{Detectability}
With the BNS and NSBH parameters summarized above, we generate the simulated signal for each GW event and apply our selection criteria $\rho_*=12$. 
Taking into account the influence of different physical effects on these two types of binary systems, we use two waveform approximants: the phenomenological model \BNSfull~\cite{Dietrich:2018uni} for BNS mergers and \NSBHfull~\cite{Thompson:2020nei} for NSBH mergers. Then, we project the waveforms onto different detectors to obtain the detected strain signal of each detector. The signal-to-noise ratio (SNR) $\rho$ is calculated as
\begin{equation}
\rho^{2}=4 \int_{f_{min}}^{f_{max}} d f \frac{|\tilde{h}(f)|^2}{S_n(f)},
\end{equation}
where $S_n(f)$ is the one-sided power spectral density of the GW detector. The considered frequency range is from $20$ to $2048$ Hz. The net SNR of the network consisting of ET+2CE detectors, as well as the root sum squared of the SNR of all detectors, is calculated using the \Bilby~\cite{Ashton:2018jfp} software package. If the net SNR exceeds the threshold value of $\rho_*=12$, the GW signal is assumed to be reliably detected and can be utilized for further analysis.
The redshift distributions of our simulated BNS and NSBH events are presented in Fig.~\ref{fig:SNR_distribution}.

\subsection{Generation of standard sirens}
\begin{figure}
\centering
\includegraphics[width=0.45\textwidth]{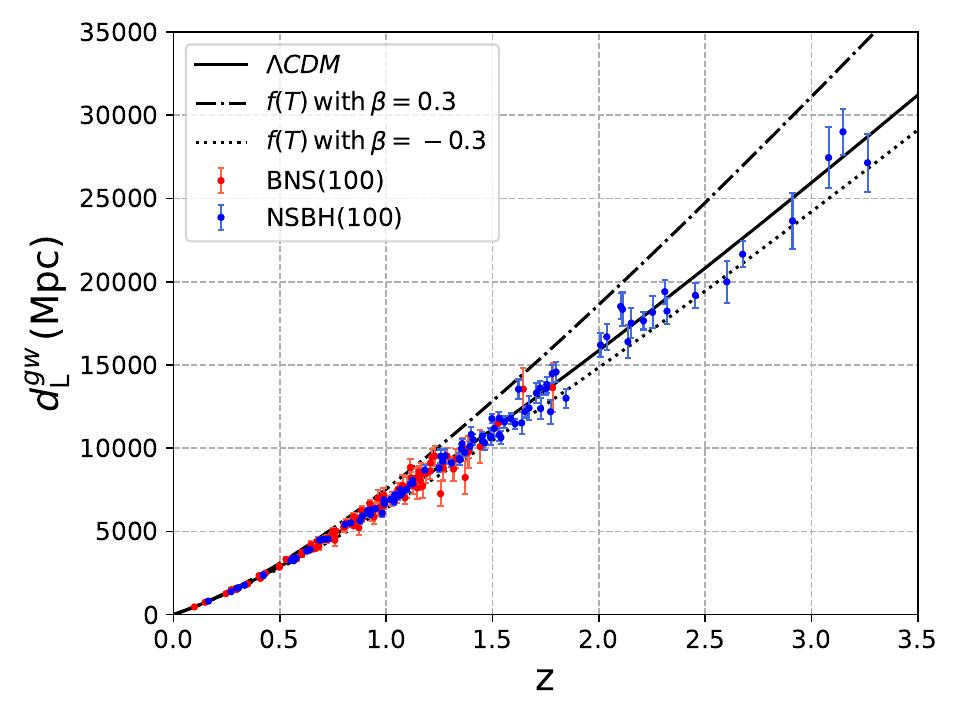}
\caption{The relation between GW luminosity distance and redshift. The solid black line represents the $\Lambda$CDM model, while the dotted and dash-dotted black lines represent the $f(T)$ model with $\beta=-0.3$ and $\beta=0.3$, respectively. The red and blue error bars represent the estimated $d_{L}^{gw}$ with 68\% credible level (\text{CL.}) for simulated BNS and NSBH events, respectively. The number of these two types of standard sirens is equal to 100.}
\label{fig:ft_z_err}
\end{figure}

The corresponding luminosity distances [$d_L\left(z_i\right)$] of the simulated BNS and NSBH events are generated based on the $\Lambda \mathrm{CDM}$ model. The values of the cosmological parameters $H_{0}$ and $\Omega_m^{(0)}$ are set as 68.6 and 0.30, respectively, as constrained by SN Ia and BAO data.

In order to expedite the calculations, we employ an analytical estimation method for $\Delta d_{\rm L}$, based on the Cutler and Flanagan approximation \cite{1994PhRvD..49.2658C}. This approximation neglects the high-mode and precession effects and has been used to estimate GW parameters in previous works~\cite{2019PhRvD.100h3514C, 2022PhRvD.106b3011W}. Since the estimated $d^{gw}_{\rm L}$ follows a Gaussian distribution $\mathcal{N}\left(d_{\rm L}\left(z_i\right),\Delta d _{{\rm L}, i}^2\right)$, we resample a new $d^{gw}_{\rm L, i, *}$ from $d^{gw}_{\rm L}$ distributions. As a result, the ``observed" luminosity distance $d_{{\rm L}, i}^{gw, \mathrm{obs}}$ follows $\mathcal{N}\left(d_{{\rm L}, i, *}\left(z_i\right),\Delta d_{{\rm L}, i}^2\right)$, providing a more realistic simulation catalog for GW events.
In previous analyses, a Gaussian likelihood has been employed to account for the uncertainties in the observed standard sirens, i.e., 
\begin{equation}
    \begin{aligned}
&\chi_{\mathrm{GW}} = \frac{d_L^{gw, \mathrm{obs}}\left(z_i\right)-d_L^{gw}\left(z_i\right)}{\sigma_i\left(z_i\right)},\\
L&=\prod_{i=1}^N \frac{1}{\sqrt{2 \pi} \sigma_i\left(z_i\right)} \exp \left(-\frac{1}{2}\chi_{\mathrm{GW}}^2\right),
    \end{aligned}
\end{equation}
where $d_L^{gw}$ is the GW ``luminosity distance" in $f(T)$ gravity, as defined by Eq.~(\ref{dl_GW}). The subscript $i$ denotes the $i$th event in the simulation catalog.

\begin{table*}[t]
	\centering
	\caption{Prior distributions of the parameters}
	\label{tab:Prior distributions}
	\setlength{\extrarowheight}{8pt}
    \begin{ruledtabular}
    \begin{tabular}{cccc}
    \multirow{1}{*}{ Name } & \multirow{1}{*}{ Parameters } & \multirow{1}{*}{Fiducial values} & \multirow{1}{*}{Priors} \\
    \hline Hubble constant & $H_0\left[\mathrm{km} \mathrm{~s}^{-1} \mathrm{Mpc}^{-1}\right]$ & $68.6$ & Uniform$(48.6,88.6)$ \\
    Matter density parameter  &$\Omega_m^{(0)}$ & $0.30$ & Uniform$(0,1)$ \\ 
    Parameter of $f(T)$ gravity &$\beta$ & $0$ & Uniform$(-0.5,0.5)$~\cite{Nesseris:2013jea,Nunes:2018evm}\textsuperscript{a} \\
    Absolute magnitude &$M$ & $-19.253$ & Uniform$(-18.253,-20.253)$~\cite{Perivolaropoulos:2022khd}\textsuperscript{a} \\
    Sound horizon &$r_{d}$ & $147.49$ & Uniform$(146.49,148.49)$~\cite{Aubourg:2014yra}\textsuperscript{a} \\
    \end{tabular}
    \end{ruledtabular}
    \footnotetext[1]{ { We have chosen prior distributions that encompass a wider range compared to the relevant references, allowing for a more comprehensive exploration of the parameter space.}}
\end{table*}

\section{RESULTS AND DISCUSSION}
\label{sec:result and discussion}
Here, we present the results of parameter estimations obtained using Monte Carlo Markov chains and discuss the potential of constraining $f(T)$ gravity using the simulated GW events. 
To obtain the posterior distributions for cosmological parameters, we employ \emcee~\cite{foreman2013emcee} as the sampler. 
Additionally, we utilize the \getdist~\cite{2019arXiv191013970L} code to facilitate the visualization of the chains and generate plots from the final results.
All of the prior distributions of the cosmological parameters are represented in \autoref{tab:Prior distributions}. Within the $\Lambda \mathrm{CDM}$ framework, fiducial values for these parameters are obtained from analyzing current observations, including SN Ia and BAOs, with best fitting values.

\begin{figure}[t]
\centering
\includegraphics[width=0.45\textwidth]{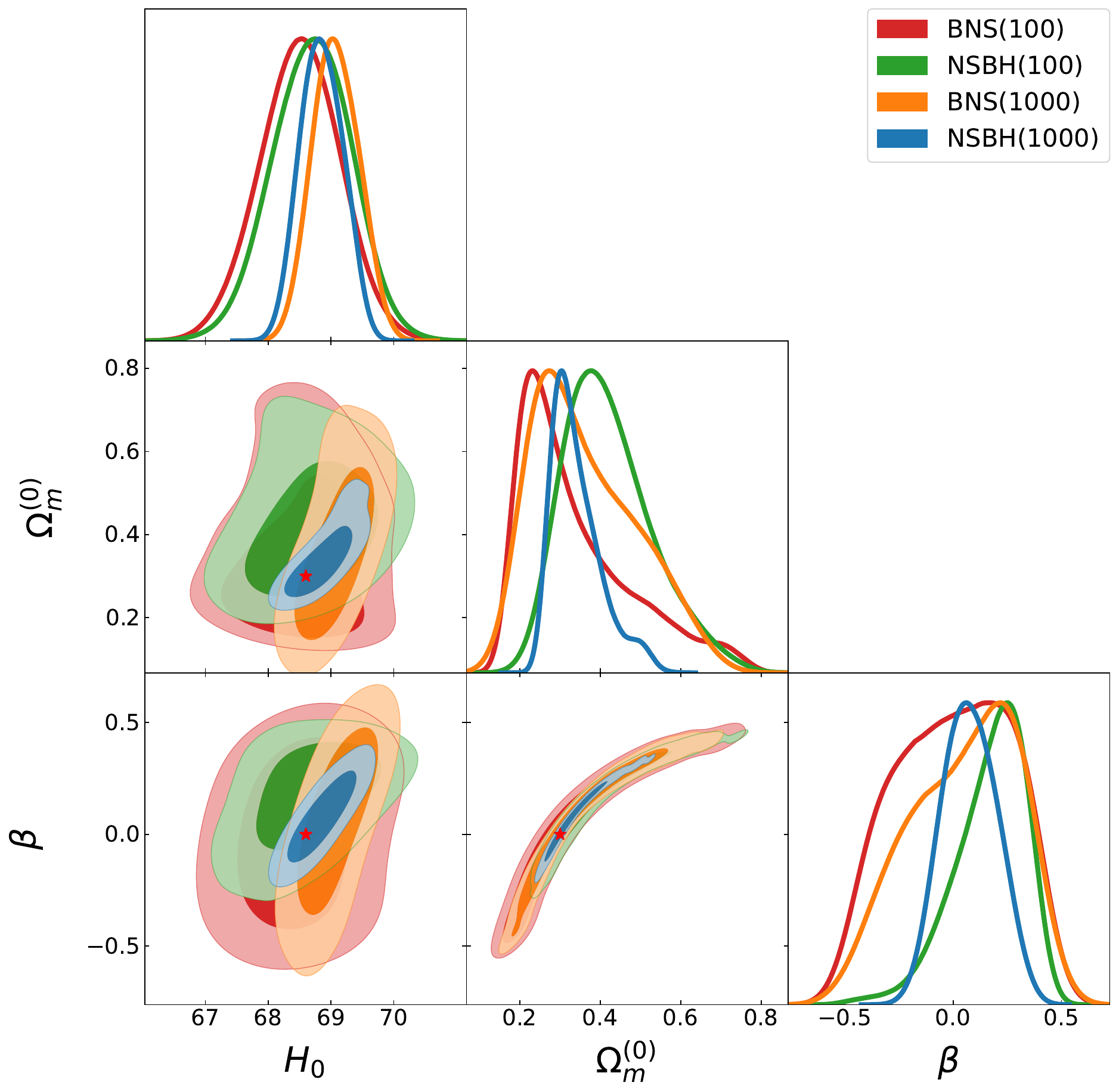}
\caption{Posterior distributions of the cosmological parameters listed in \autoref{tab:posterior distributions} utilizing GW events only. Using 100 (1000) BNS and NSBH events, the results of the power-low model of $f(T)$ gravity are marked in red, green, orange, and blue, respectively. Red stars represent the fiducial values $\left(H_{0}, \Omega_m^{(0)}, \beta\right)=(68.6,0.30,0)$. The contours are at the 68\% and 95\% \text{CL}.}

\label{fig:BNS_NSBH}
\end{figure}

\begin{figure*}[t]
\centering
\includegraphics[width=0.8\textwidth]{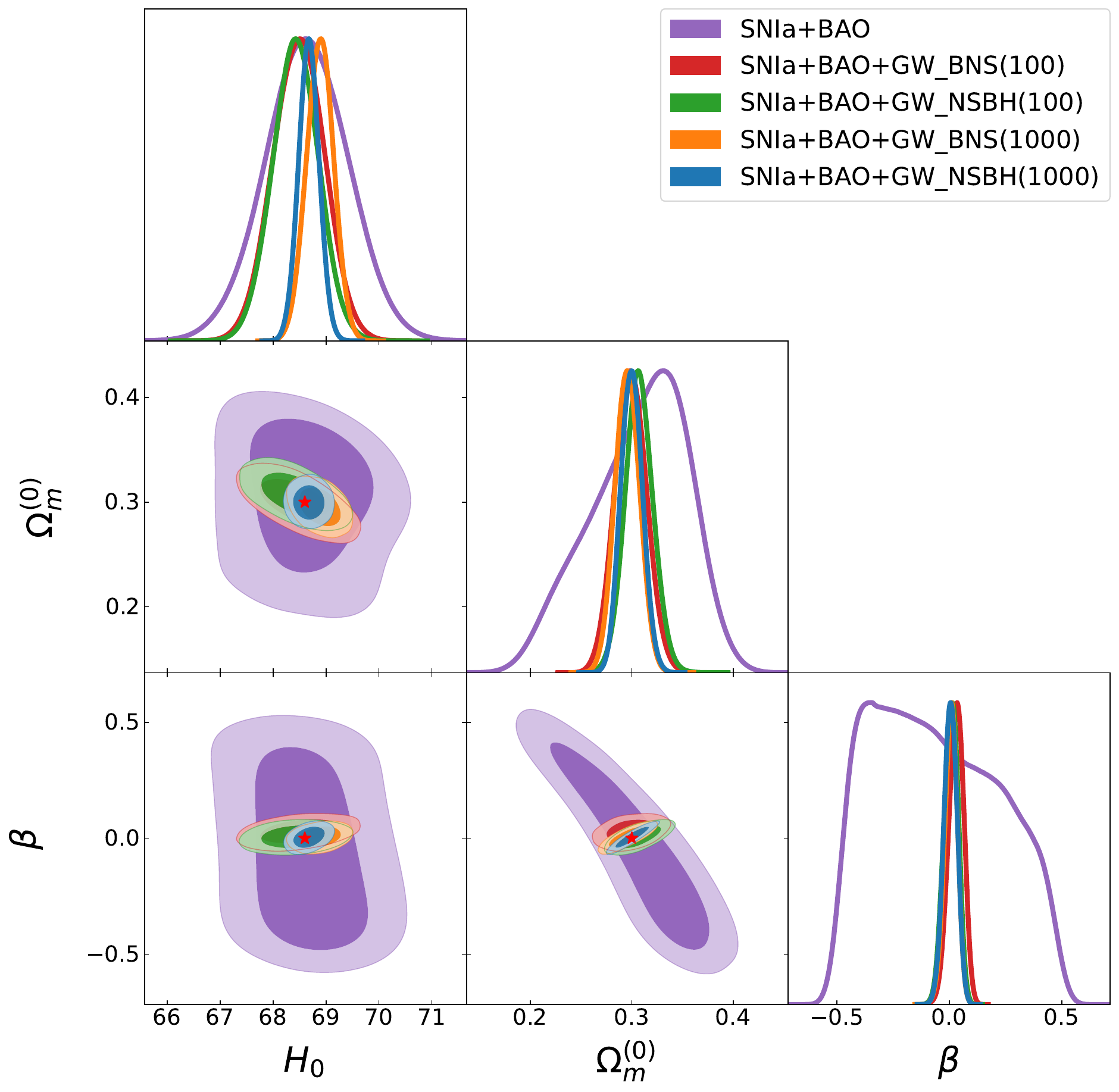}
\caption{Posterior distributions of the parameters listed in \autoref{tab:posterior distributions} utilizing GW events, SN Ia, and BAO observations. The results for the power-law model of $f(T)$ gravity, obtained using different combinations of data, are represented by different colors. Red stars represent the fiducial values $\left(H_{0}, \Omega_m^{(0)}, \beta\right)=(68.6,0.30,0)$. The contours are at the 68\% and 95\% \text{CL}.}
\label{fig:SNIa_BAO_BNS_NSBH}
\end{figure*}

As shown in Fig.~\ref{fig:ft_z_err}, the discrepancies between $d_{\rm L}^{gw}$ and $d_{\rm L}^{\mathrm{em}}$ caused by parameters changing under $f(T)$ gravity become remarkable at higher redshifts. Consequently, the NSBH samples are anticipated to offer more stringent constraints on $f(T)$ gravity compared to the BNS samples, as illustrated in Fig.~\ref{fig:BNS_NSBH}. Moreover, there is a substantial degeneracy between the parameters of $\Omega_{m}^{(0)}$ and $\beta$ for analysis of both binary systems. 
This degeneracy arises from  the combined term of $\beta$ and $\Omega_{m}^{(0)}$ in Eqs.~(\ref{eq:efun}) and (\ref{delta}). The former equation plays a significant role in the analysis of SN Ia and BAO data and the latter one dominates the analysis of GWs, generating different degeneracy in Figs.~\ref{fig:BNS_NSBH} and \ref{fig:SNIa_BAO_BNS_NSBH}.

\begin{table*}[t] 
	\centering
	\caption{Posterior results\textsuperscript{a} of the parameters for $f(T)$ gravity}
	\label{tab:posterior distributions}
	\setlength{\extrarowheight}{8pt}
    \begin{ruledtabular}
    \begin{tabular}{cccccc}
    \multirow{1}{*}{ Data sets\textsuperscript{b} } & \multirow{1}{*}{$H_0\left[\mathrm{~km} \mathrm{~s}^{-1} \mathrm{Mpc}^{-1}\right]$} & \multirow{1}{*}{$\Omega_{m}$} & \multirow{1}{*}{$\beta$} & \multirow{1}{*}{$M$} & \multirow{1}{*}{$r_{d}$}\\
    \hline $\mathrm{SN Ia}+\mathrm{BAO}(\Lambda \rm CDM)$ & $68.6_{-0.8}^{+0.7}$ & $0.300_{-0.020}^{+0.024}$ & $-$ & $-19.392_{-0.021}^{+0.019}$ &$146.85_{-0.04}^{+1.32}$\\
    \hline
    $\mathrm{SN Ia}+\mathrm{BAO}$ & $68.6_{-0.8}^{+0.8}$ & $0.331_{-0.061}^{+0.034}$ & $-0.35_{-0.10}^{+0.49}$ & $-19.390_{-0.020}^{+0.019}$ &$146.76_{-0.05}^{+1.42}$\\
    $\mathrm{BNS(100)}$ & $68.5_{-0.7}^{+0.6}$ & $0.231_{-0.064}^{+0.165}$ & $0.17_{-0.41}^{+0.16}$ & $-$ & $-$\\
    $\mathrm{BNS(1000)}$ & $69.0_{-0.3}^{+0.4}$ & $0.273_{-0.079}^{+0.183}$ & $0.22_{-0.38}^{+0.15}$ & $-$ & $-$\\
    $\mathrm{NSBH(100)}$ & $68.7_{-0.7}^{+0.6}$ & $0.376_{-0.089}^{+0.124}$ & $0.25_{-0.33}^{+0.28}$ & $-$ & $-$ \\
    $\mathrm{NSBH(1000)}$ & $68.8_{-0.3}^{+0.4}$ & $0.303_{-0.038}^{+0.071}$ & $0.06_{-0.12}^{+0.14}$ & $-$ & $-$ \\
    $\mathrm{SN Ia}+\mathrm{BAO}+\mathrm{BNS(100)}$ & $68.5_{-0.5}^{+0.5}$ & $0.298_{-0.015}^{+0.016}$ & $0.034_{-0.035}^{+0.031}$ & $-19.394_{-0.015}^{+0.014}$ &
    $148.11_{-1.61}^{+0.38}$\\
    
    $\mathrm{SN Ia}+\mathrm{BAO}+\mathrm{BNS(1000)}$ & $68.9_{-0.3}^{+0.2}$ & $0.296_{-0.012}^{+0.012}$ & $0.009_{-0.030}^{+0.027}$ & $-19.385_{-0.011}^{+0.010}$ &
    $146.67_{-0.18}^{+1.82}$\\
    
    $\mathrm{SN Ia}+\mathrm{BAO}+\mathrm{NSBH(100)}$ & $68.4_{-0.4}^{+0.4}$ & $0.306_{-0.013}^{+0.015}$ & $0.012_{-0.035}^{+0.029}$ & $-19.394_{-0.014}^{+0.014}$ &
    $147.91_{-1.42}^{+0.58}$ \\
    
    $\mathrm{SN Ia}+\mathrm{BAO}+\mathrm{NSBH(1000)}$ & $68.7_{-0.2}^{+0.2}$ & $0.300_{-0.011}^{+0.011}$ & $0.005_{-0.030}^{+0.028}$ & $-19.390_{-0.010}^{+0.011}$ &
    $147.36_{-0.71}^{+0.58}$ \\

    \end{tabular}
    \end{ruledtabular}
    \footnotetext[1]{ The uncertainties are computed based on Highest probability distribution (HPD) of posterior distributions}

    \footnotetext[2]{ (Number) represents the number of events }

\end{table*}

The results obtained from combining GW simulations, SN Ia, and BAO observations are shown in \autoref{tab:posterior distributions} and Fig.~\ref{fig:SNIa_BAO_BNS_NSBH}.
Compared to the analysis of SN Ia and BAO observations alone, which yielded $\Omega_m^{(0)}=0.331_{-0.061}^{+0.034}$ and $\beta=-0.35_{-0.10}^{+0.49}$ at 68\% \text{CL.}, the combined analysis breaks the degeneracy between these two parameters and provides more precise estimates for them.
Specifically, the combined analysis from SN Ia + BAO + NSBH (1000) yields $\Omega_m^{(0)}=0.300_{-0.011}^{+0.011}$ and $\beta=0.005_{-0.030}^{+0.028}$ at 68\% \text{CL}.

This improvement can also be demonstrated through comparisons of the uncertainties in $H_0$ obtained from different datasets,
\begin{equation}
    \frac{\Delta H_0^{c}}{\Delta H_0^{a}}=9.8 \%, \quad \frac{\Delta H_0^{c}}{\Delta H_0^{b}}=22.3 \%,
\end{equation}
where $\Delta H_0^{a}$, $\Delta H_0^{b}$, and$\Delta H_0^{c}$ represent the uncertainties of $H_{0}$ estimation within the 68\% \text{CL.} for different datasets: SN Ia + BAO, NSBH (1000) and SN Ia + BAO + NSBH (1000), respectively. In Fig.~\ref{fig:SNIa_BAO_BNS_NSBH}, we also compare the estimated precision between BNS and NSBH events.
Overall, after incorporating high-redshift GW events with available SN Ia and BAO observations, the precise constraints on $H_0$ and $f(T)$ gravity can be achieved, thereby enhancing our understanding of gravity and cosmology in the future.

\section{Conclusion}
\label{sec:Conclusion}
The developing field of multimessenger astronomy, facilitated by GWs, provides a novel avenue for investigating gravity and cosmology.
In this work, we focus on utilizing the simulated BNS and NSBH events to prospectively constrain $f(T)$ gravity with the third-generation gravitational-wave detectors. 

Based on current observations, we conducted a comprehensive simulation of the BNS and NSBH mergers, taking into account realistic selection criteria to ensure that these simulated events are always accompanied with electromagnetic counterparts with certain redshifts. Moreover, we restrict the inclination angle to a small range ($\sim 0.1$ $\mathrm{rad}$) to enhance the realism and precision of uncertainty estimation in the GW luminosity distance.
The expected deviations in the ratio $d_L^{gw} / d_L^{em}$ caused by $f(T)$ gravity are anticipated to exceed 15\% at high redshifts ($z \sim 5$). Therefore, we can compare the modified $d_{\rm L}^{gw}$ with $d_{\rm L}^{em}$ to constrain $f(T)$ gravity and other cosmological parameters.

By analyzing SN Ia and BAO data, we established the baseline of simulated GW events in the frame of $\Lambda \mathrm{CDM}$ and constrained the parameter $\beta=-0.35_{-0.10}^{+0.49}$ at 68\% \text{CL.} in the power-law model under $f(T)$ gravity.
Moreover, we incorporated simulated GW events into our analysis, combined with SN Ia and BAO observations, and discovered that this combination effectively breaks the degeneracy between $\Omega_m^{(0)}$ and $\beta$, leading to the tightest constraints that $\beta = 0.005_{-0.030}^{+0.028}$ at 68\% \text{CL}.
This enhanced precision shows an order of magnitude improvement over the analyzing result based on SN Ia and BAO data, which enables us to determine the deviation between modified gravity and GR.
In addition, the GW simulated events provide $H_0$ with uncertainties ranging from 1.0\% to 1.9\%, indicating a promising avenue to explore the $H_0$ tension under the $f(T)$ gravity framework.

In summary, the third-generation gravitational-wave detectors provide a valuable opportunity to impose more stringent constraints on $f(T)$ gravity, contributing significantly to our improved understanding of gravity and cosmology.

\begin{acknowledgments}
We are grateful to Wen Zhao, ShaoPeng Tang, MingZhe Han, Rui Niu, Zhao Li, Jin Qiao, Xin Ren, and BinBin Zhang for helpful discussions.
This work is supported in part by the National Natural Science Foundation of China (No. 12233011 and 11921003).
\end{acknowledgments}

\bibliographystyle{apsrev4-2}
\bibliography{ref.bib}

\providecommand{\noopsort}[1]{}\providecommand{\singleletter}[1]{#1}%
\begin{thebibliography}{80}%
\makeatletter
\providecommand \@ifxundefined [1]{%
 \@ifx{#1\undefined}
}%
\providecommand \@ifnum [1]{%
 \ifnum #1\expandafter \@firstoftwo
 \else \expandafter \@secondoftwo
 \fi
}%
\providecommand \@ifx [1]{%
 \ifx #1\expandafter \@firstoftwo
 \else \expandafter \@secondoftwo
 \fi
}%
\providecommand \natexlab [1]{#1}%
\providecommand \enquote  [1]{``#1''}%
\providecommand \bibnamefont  [1]{#1}%
\providecommand \bibfnamefont [1]{#1}%
\providecommand \citenamefont [1]{#1}%
\providecommand \href@noop [0]{\@secondoftwo}%
\providecommand \href [0]{\begingroup \@sanitize@url \@href}%
\providecommand \@href[1]{\@@startlink{#1}\@@href}%
\providecommand \@@href[1]{\endgroup#1\@@endlink}%
\providecommand \@sanitize@url [0]{\catcode `\\12\catcode `\$12\catcode
  `\&12\catcode `\#12\catcode `\^12\catcode `\_12\catcode `\%12\relax}%
\providecommand \@@startlink[1]{}%
\providecommand \@@endlink[0]{}%
\providecommand \url  [0]{\begingroup\@sanitize@url \@url }%
\providecommand \@url [1]{\endgroup\@href {#1}{\urlprefix }}%
\providecommand \urlprefix  [0]{URL }%
\providecommand \Eprint [0]{\href }%
\providecommand \doibase [0]{https://doi.org/}%
\providecommand \selectlanguage [0]{\@gobble}%
\providecommand \bibinfo  [0]{\@secondoftwo}%
\providecommand \bibfield  [0]{\@secondoftwo}%
\providecommand \translation [1]{[#1]}%
\providecommand \BibitemOpen [0]{}%
\providecommand \bibitemStop [0]{}%
\providecommand \bibitemNoStop [0]{.\EOS\space}%
\providecommand \EOS [0]{\spacefactor3000\relax}%
\providecommand \BibitemShut  [1]{\csname bibitem#1\endcsname}%
\let\auto@bib@innerbib\@empty
\bibitem [{\citenamefont {Abbott}\ \emph {et~al.}(2016)\citenamefont {Abbott}
  \emph {et~al.}}]{LIGOScientific:2016aoc}%
  \BibitemOpen
  \bibfield  {author} {\bibinfo {author} {\bibfnamefont {B.~P.}\ \bibnamefont
  {Abbott}} \emph {et~al.} (\bibinfo {collaboration} {LIGO Scientific,
  Virgo}),\ }\href {https://doi.org/10.1103/PhysRevLett.116.061102} {\bibfield
  {journal} {\bibinfo  {journal} {Phys. Rev. Lett.}\ }\textbf {\bibinfo
  {volume} {116}},\ \bibinfo {pages} {061102} (\bibinfo {year} {2016})},\
  \Eprint {https://arxiv.org/abs/1602.03837} {arXiv:1602.03837 [gr-qc]}
  \BibitemShut {NoStop}%
\bibitem [{\citenamefont {Abbott}\ \emph
  {et~al.}(2017{\natexlab{a}})\citenamefont {Abbott} \emph
  {et~al.}}]{LIGOScientific:2017vwq}%
  \BibitemOpen
  \bibfield  {author} {\bibinfo {author} {\bibfnamefont {B.~P.}\ \bibnamefont
  {Abbott}} \emph {et~al.} (\bibinfo {collaboration} {LIGO Scientific,
  Virgo}),\ }\href {https://doi.org/10.1103/PhysRevLett.119.161101} {\bibfield
  {journal} {\bibinfo  {journal} {Phys. Rev. Lett.}\ }\textbf {\bibinfo
  {volume} {119}},\ \bibinfo {pages} {161101} (\bibinfo {year}
  {2017}{\natexlab{a}})},\ \Eprint {https://arxiv.org/abs/1710.05832}
  {arXiv:1710.05832 [gr-qc]} \BibitemShut {NoStop}%
\bibitem [{\citenamefont {Shankaranarayanan}\ and\ \citenamefont
  {Johnson}(2022)}]{Shankaranarayanan:2022wbx}%
  \BibitemOpen
  \bibfield  {author} {\bibinfo {author} {\bibfnamefont {S.}~\bibnamefont
  {Shankaranarayanan}}\ and\ \bibinfo {author} {\bibfnamefont {J.~P.}\
  \bibnamefont {Johnson}},\ }\href {https://doi.org/10.1007/s10714-022-02927-2}
  {\bibfield  {journal} {\bibinfo  {journal} {Gen. Rel. Grav.}\ }\textbf
  {\bibinfo {volume} {54}},\ \bibinfo {pages} {44} (\bibinfo {year} {2022})},\
  \Eprint {https://arxiv.org/abs/2204.06533} {arXiv:2204.06533 [gr-qc]}
  \BibitemShut {NoStop}%
\bibitem [{\citenamefont {Clifton}\ \emph {et~al.}(2012)\citenamefont
  {Clifton}, \citenamefont {Ferreira}, \citenamefont {Padilla},\ and\
  \citenamefont {Skordis}}]{Clifton:2011jh}%
  \BibitemOpen
  \bibfield  {author} {\bibinfo {author} {\bibfnamefont {T.}~\bibnamefont
  {Clifton}}, \bibinfo {author} {\bibfnamefont {P.~G.}\ \bibnamefont
  {Ferreira}}, \bibinfo {author} {\bibfnamefont {A.}~\bibnamefont {Padilla}},\
  and\ \bibinfo {author} {\bibfnamefont {C.}~\bibnamefont {Skordis}},\ }\href
  {https://doi.org/10.1016/j.physrep.2012.01.001} {\bibfield  {journal}
  {\bibinfo  {journal} {Phys. Rept.}\ }\textbf {\bibinfo {volume} {513}},\
  \bibinfo {pages} {1} (\bibinfo {year} {2012})},\ \Eprint
  {https://arxiv.org/abs/1106.2476} {arXiv:1106.2476 [astro-ph.CO]}
  \BibitemShut {NoStop}%
\bibitem [{\citenamefont {Dolgov}\ and\ \citenamefont
  {Kawasaki}(2003)}]{Dolgov:2003px}%
  \BibitemOpen
  \bibfield  {author} {\bibinfo {author} {\bibfnamefont {A.~D.}\ \bibnamefont
  {Dolgov}}\ and\ \bibinfo {author} {\bibfnamefont {M.}~\bibnamefont
  {Kawasaki}},\ }\href {https://doi.org/10.1016/j.physletb.2003.08.039}
  {\bibfield  {journal} {\bibinfo  {journal} {Phys. Lett. B}\ }\textbf
  {\bibinfo {volume} {573}},\ \bibinfo {pages} {1} (\bibinfo {year} {2003})},\
  \Eprint {https://arxiv.org/abs/astro-ph/0307285} {arXiv:astro-ph/0307285}
  \BibitemShut {NoStop}%
\bibitem [{\citenamefont {Sotiriou}\ and\ \citenamefont
  {Faraoni}(2010)}]{Sotiriou:2008rp}%
  \BibitemOpen
  \bibfield  {author} {\bibinfo {author} {\bibfnamefont {T.~P.}\ \bibnamefont
  {Sotiriou}}\ and\ \bibinfo {author} {\bibfnamefont {V.}~\bibnamefont
  {Faraoni}},\ }\href {https://doi.org/10.1103/RevModPhys.82.451} {\bibfield
  {journal} {\bibinfo  {journal} {Rev. Mod. Phys.}\ }\textbf {\bibinfo {volume}
  {82}},\ \bibinfo {pages} {451} (\bibinfo {year} {2010})},\ \Eprint
  {https://arxiv.org/abs/0805.1726} {arXiv:0805.1726 [gr-qc]} \BibitemShut
  {NoStop}%
\bibitem [{\citenamefont {De~Felice}\ and\ \citenamefont
  {Tsujikawa}(2010)}]{DeFelice:2010aj}%
  \BibitemOpen
  \bibfield  {author} {\bibinfo {author} {\bibfnamefont {A.}~\bibnamefont
  {De~Felice}}\ and\ \bibinfo {author} {\bibfnamefont {S.}~\bibnamefont
  {Tsujikawa}},\ }\href {https://doi.org/10.12942/lrr-2010-3} {\bibfield
  {journal} {\bibinfo  {journal} {Living Rev. Rel.}\ }\textbf {\bibinfo
  {volume} {13}},\ \bibinfo {pages} {3} (\bibinfo {year} {2010})},\ \Eprint
  {https://arxiv.org/abs/1002.4928} {arXiv:1002.4928 [gr-qc]} \BibitemShut
  {NoStop}%
\bibitem [{\citenamefont {Maluf}(2013)}]{Maluf:2013gaa}%
  \BibitemOpen
  \bibfield  {author} {\bibinfo {author} {\bibfnamefont {J.~W.}\ \bibnamefont
  {Maluf}},\ }\href {https://doi.org/10.1002/andp.201200272} {\bibfield
  {journal} {\bibinfo  {journal} {Annalen Phys.}\ }\textbf {\bibinfo {volume}
  {525}},\ \bibinfo {pages} {339} (\bibinfo {year} {2013})},\ \Eprint
  {https://arxiv.org/abs/1303.3897} {arXiv:1303.3897 [gr-qc]} \BibitemShut
  {NoStop}%
\bibitem [{\citenamefont {Cai}\ \emph {et~al.}(2016)\citenamefont {Cai},
  \citenamefont {Capozziello}, \citenamefont {De~Laurentis},\ and\
  \citenamefont {Saridakis}}]{Cai:2015emx}%
  \BibitemOpen
  \bibfield  {author} {\bibinfo {author} {\bibfnamefont {Y.-F.}\ \bibnamefont
  {Cai}}, \bibinfo {author} {\bibfnamefont {S.}~\bibnamefont {Capozziello}},
  \bibinfo {author} {\bibfnamefont {M.}~\bibnamefont {De~Laurentis}},\ and\
  \bibinfo {author} {\bibfnamefont {E.~N.}\ \bibnamefont {Saridakis}},\ }\href
  {https://doi.org/10.1088/0034-4885/79/10/106901} {\bibfield  {journal}
  {\bibinfo  {journal} {Rept. Prog. Phys.}\ }\textbf {\bibinfo {volume} {79}},\
  \bibinfo {pages} {106901} (\bibinfo {year} {2016})},\ \Eprint
  {https://arxiv.org/abs/1511.07586} {arXiv:1511.07586 [gr-qc]} \BibitemShut
  {NoStop}%
\bibitem [{\citenamefont {Bahamonde}\ \emph {et~al.}(2023)\citenamefont
  {Bahamonde}, \citenamefont {Dialektopoulos}, \citenamefont
  {Escamilla-Rivera}, \citenamefont {Farrugia}, \citenamefont {Gakis},
  \citenamefont {Hendry}, \citenamefont {Hohmann}, \citenamefont {Levi~Said},
  \citenamefont {Mifsud},\ and\ \citenamefont
  {Di~Valentino}}]{Bahamonde:2021gfp}%
  \BibitemOpen
  \bibfield  {author} {\bibinfo {author} {\bibfnamefont {S.}~\bibnamefont
  {Bahamonde}}, \bibinfo {author} {\bibfnamefont {K.~F.}\ \bibnamefont
  {Dialektopoulos}}, \bibinfo {author} {\bibfnamefont {C.}~\bibnamefont
  {Escamilla-Rivera}}, \bibinfo {author} {\bibfnamefont {G.}~\bibnamefont
  {Farrugia}}, \bibinfo {author} {\bibfnamefont {V.}~\bibnamefont {Gakis}},
  \bibinfo {author} {\bibfnamefont {M.}~\bibnamefont {Hendry}}, \bibinfo
  {author} {\bibfnamefont {M.}~\bibnamefont {Hohmann}}, \bibinfo {author}
  {\bibfnamefont {J.}~\bibnamefont {Levi~Said}}, \bibinfo {author}
  {\bibfnamefont {J.}~\bibnamefont {Mifsud}},\ and\ \bibinfo {author}
  {\bibfnamefont {E.}~\bibnamefont {Di~Valentino}},\ }\href
  {https://doi.org/10.1088/1361-6633/ac9cef} {\bibfield  {journal} {\bibinfo
  {journal} {Rept. Prog. Phys.}\ }\textbf {\bibinfo {volume} {86}},\ \bibinfo
  {pages} {026901} (\bibinfo {year} {2023})},\ \Eprint
  {https://arxiv.org/abs/2106.13793} {arXiv:2106.13793 [gr-qc]} \BibitemShut
  {NoStop}%
\bibitem [{\citenamefont {Krssak}\ \emph {et~al.}(2019)\citenamefont {Krssak},
  \citenamefont {van~den Hoogen}, \citenamefont {Pereira}, \citenamefont
  {B\"ohmer},\ and\ \citenamefont {Coley}}]{Krssak:2018ywd}%
  \BibitemOpen
  \bibfield  {author} {\bibinfo {author} {\bibfnamefont {M.}~\bibnamefont
  {Krssak}}, \bibinfo {author} {\bibfnamefont {R.~J.}\ \bibnamefont {van~den
  Hoogen}}, \bibinfo {author} {\bibfnamefont {J.~G.}\ \bibnamefont {Pereira}},
  \bibinfo {author} {\bibfnamefont {C.~G.}\ \bibnamefont {B\"ohmer}},\ and\
  \bibinfo {author} {\bibfnamefont {A.~A.}\ \bibnamefont {Coley}},\ }\href
  {https://doi.org/10.1088/1361-6382/ab2e1f} {\bibfield  {journal} {\bibinfo
  {journal} {Class. Quant. Grav.}\ }\textbf {\bibinfo {volume} {36}},\ \bibinfo
  {pages} {183001} (\bibinfo {year} {2019})},\ \Eprint
  {https://arxiv.org/abs/1810.12932} {arXiv:1810.12932 [gr-qc]} \BibitemShut
  {NoStop}%
\bibitem [{\citenamefont {Bahamonde}\ \emph {et~al.}(2015)\citenamefont
  {Bahamonde}, \citenamefont {B\"ohmer},\ and\ \citenamefont
  {Wright}}]{Bahamonde:2015zma}%
  \BibitemOpen
  \bibfield  {author} {\bibinfo {author} {\bibfnamefont {S.}~\bibnamefont
  {Bahamonde}}, \bibinfo {author} {\bibfnamefont {C.~G.}\ \bibnamefont
  {B\"ohmer}},\ and\ \bibinfo {author} {\bibfnamefont {M.}~\bibnamefont
  {Wright}},\ }\href {https://doi.org/10.1103/PhysRevD.92.104042} {\bibfield
  {journal} {\bibinfo  {journal} {Phys. Rev. D}\ }\textbf {\bibinfo {volume}
  {92}},\ \bibinfo {pages} {104042} (\bibinfo {year} {2015})},\ \Eprint
  {https://arxiv.org/abs/1508.05120} {arXiv:1508.05120 [gr-qc]} \BibitemShut
  {NoStop}%
\bibitem [{\citenamefont {Cai}\ \emph {et~al.}(2018)\citenamefont {Cai},
  \citenamefont {Li}, \citenamefont {Saridakis},\ and\ \citenamefont
  {Xue}}]{Cai:2018rzd}%
  \BibitemOpen
  \bibfield  {author} {\bibinfo {author} {\bibfnamefont {Y.-F.}\ \bibnamefont
  {Cai}}, \bibinfo {author} {\bibfnamefont {C.}~\bibnamefont {Li}}, \bibinfo
  {author} {\bibfnamefont {E.~N.}\ \bibnamefont {Saridakis}},\ and\ \bibinfo
  {author} {\bibfnamefont {L.}~\bibnamefont {Xue}},\ }\href
  {https://doi.org/10.1103/PhysRevD.97.103513} {\bibfield  {journal} {\bibinfo
  {journal} {Phys. Rev. D}\ }\textbf {\bibinfo {volume} {97}},\ \bibinfo
  {pages} {103513} (\bibinfo {year} {2018})},\ \Eprint
  {https://arxiv.org/abs/1801.05827} {arXiv:1801.05827 [gr-qc]} \BibitemShut
  {NoStop}%
\bibitem [{\citenamefont {Nunes}\ \emph {et~al.}(2018)\citenamefont {Nunes},
  \citenamefont {Pan},\ and\ \citenamefont {Saridakis}}]{Nunes:2018evm}%
  \BibitemOpen
  \bibfield  {author} {\bibinfo {author} {\bibfnamefont {R.~C.}\ \bibnamefont
  {Nunes}}, \bibinfo {author} {\bibfnamefont {S.}~\bibnamefont {Pan}},\ and\
  \bibinfo {author} {\bibfnamefont {E.~N.}\ \bibnamefont {Saridakis}},\ }\href
  {https://doi.org/10.1103/PhysRevD.98.104055} {\bibfield  {journal} {\bibinfo
  {journal} {Phys. Rev. D}\ }\textbf {\bibinfo {volume} {98}},\ \bibinfo
  {pages} {104055} (\bibinfo {year} {2018})},\ \Eprint
  {https://arxiv.org/abs/1810.03942} {arXiv:1810.03942 [gr-qc]} \BibitemShut
  {NoStop}%
\bibitem [{\citenamefont {Kumar}\ \emph {et~al.}(2023)\citenamefont {Kumar},
  \citenamefont {Nunes},\ and\ \citenamefont {Yadav}}]{Kumar:2022nvf}%
  \BibitemOpen
  \bibfield  {author} {\bibinfo {author} {\bibfnamefont {S.}~\bibnamefont
  {Kumar}}, \bibinfo {author} {\bibfnamefont {R.~C.}\ \bibnamefont {Nunes}},\
  and\ \bibinfo {author} {\bibfnamefont {P.}~\bibnamefont {Yadav}},\ }\href
  {https://doi.org/10.1103/PhysRevD.107.063529} {\bibfield  {journal} {\bibinfo
   {journal} {Phys. Rev. D}\ }\textbf {\bibinfo {volume} {107}},\ \bibinfo
  {pages} {063529} (\bibinfo {year} {2023})},\ \Eprint
  {https://arxiv.org/abs/2209.11131} {arXiv:2209.11131 [astro-ph.CO]}
  \BibitemShut {NoStop}%
\bibitem [{\citenamefont {Nunes}\ \emph {et~al.}(2019)\citenamefont {Nunes},
  \citenamefont {Alves},\ and\ \citenamefont {de~Araujo}}]{Nunes:2019bjq}%
  \BibitemOpen
  \bibfield  {author} {\bibinfo {author} {\bibfnamefont {R.~C.}\ \bibnamefont
  {Nunes}}, \bibinfo {author} {\bibfnamefont {M.~E.~S.}\ \bibnamefont
  {Alves}},\ and\ \bibinfo {author} {\bibfnamefont {J.~C.~N.}\ \bibnamefont
  {de~Araujo}},\ }\href {https://doi.org/10.1103/PhysRevD.100.064012}
  {\bibfield  {journal} {\bibinfo  {journal} {Phys. Rev. D}\ }\textbf {\bibinfo
  {volume} {100}},\ \bibinfo {pages} {064012} (\bibinfo {year} {2019})},\
  \Eprint {https://arxiv.org/abs/1905.03237} {arXiv:1905.03237 [gr-qc]}
  \BibitemShut {NoStop}%
\bibitem [{\citenamefont {Zhang}\ and\ \citenamefont
  {Zhang}(2021)}]{Zhang:2021kqn}%
  \BibitemOpen
  \bibfield  {author} {\bibinfo {author} {\bibfnamefont {Y.}~\bibnamefont
  {Zhang}}\ and\ \bibinfo {author} {\bibfnamefont {H.}~\bibnamefont {Zhang}},\
  }\href {https://doi.org/10.1140/epjc/s10052-021-09501-1} {\bibfield
  {journal} {\bibinfo  {journal} {Eur. Phys. J. C}\ }\textbf {\bibinfo {volume}
  {81}},\ \bibinfo {pages} {706} (\bibinfo {year} {2021})},\ \Eprint
  {https://arxiv.org/abs/2108.05736} {arXiv:2108.05736 [astro-ph.CO]}
  \BibitemShut {NoStop}%
\bibitem [{\citenamefont {Tzerefos}\ \emph {et~al.}(2023)\citenamefont
  {Tzerefos}, \citenamefont {Papanikolaou}, \citenamefont {Saridakis},\ and\
  \citenamefont {Basilakos}}]{Tzerefos:2023mpe}%
  \BibitemOpen
  \bibfield  {author} {\bibinfo {author} {\bibfnamefont {C.}~\bibnamefont
  {Tzerefos}}, \bibinfo {author} {\bibfnamefont {T.}~\bibnamefont
  {Papanikolaou}}, \bibinfo {author} {\bibfnamefont {E.~N.}\ \bibnamefont
  {Saridakis}},\ and\ \bibinfo {author} {\bibfnamefont {S.}~\bibnamefont
  {Basilakos}},\ }\href {https://doi.org/10.1103/PhysRevD.107.124019}
  {\bibfield  {journal} {\bibinfo  {journal} {Phys. Rev. D}\ }\textbf {\bibinfo
  {volume} {107}},\ \bibinfo {pages} {124019} (\bibinfo {year} {2023})},\
  \Eprint {https://arxiv.org/abs/2303.16695} {arXiv:2303.16695 [gr-qc]}
  \BibitemShut {NoStop}%
\bibitem [{\citenamefont {Briffa}\ \emph {et~al.}(2023)\citenamefont {Briffa},
  \citenamefont {Escamilla-Rivera}, \citenamefont {Levi~Said},\ and\
  \citenamefont {Mifsud}}]{Briffa:2023ern}%
  \BibitemOpen
  \bibfield  {author} {\bibinfo {author} {\bibfnamefont {R.}~\bibnamefont
  {Briffa}}, \bibinfo {author} {\bibfnamefont {C.}~\bibnamefont
  {Escamilla-Rivera}}, \bibinfo {author} {\bibfnamefont {J.}~\bibnamefont
  {Levi~Said}},\ and\ \bibinfo {author} {\bibfnamefont {J.}~\bibnamefont
  {Mifsud}},\ }\href {https://doi.org/10.1093/mnras/stad1384} {\bibfield
  {journal} {\bibinfo  {journal} {Mon. Not. Roy. Astron. Soc.}\ }\textbf
  {\bibinfo {volume} {522}},\ \bibinfo {pages} {6024} (\bibinfo {year}
  {2023})},\ \Eprint {https://arxiv.org/abs/2303.13840} {arXiv:2303.13840
  [gr-qc]} \BibitemShut {NoStop}%
\bibitem [{\citenamefont {Nishizawa}(2018)}]{Nishizawa:2017nef}%
  \BibitemOpen
  \bibfield  {author} {\bibinfo {author} {\bibfnamefont {A.}~\bibnamefont
  {Nishizawa}},\ }\href {https://doi.org/10.1103/PhysRevD.97.104037} {\bibfield
   {journal} {\bibinfo  {journal} {Phys. Rev. D}\ }\textbf {\bibinfo {volume}
  {97}},\ \bibinfo {pages} {104037} (\bibinfo {year} {2018})},\ \Eprint
  {https://arxiv.org/abs/1710.04825} {arXiv:1710.04825 [gr-qc]} \BibitemShut
  {NoStop}%
\bibitem [{\citenamefont {Abbott}\ \emph
  {et~al.}(2017{\natexlab{b}})\citenamefont {Abbott} \emph
  {et~al.}}]{LIGOScientific:2017zic}%
  \BibitemOpen
  \bibfield  {author} {\bibinfo {author} {\bibfnamefont {B.~P.}\ \bibnamefont
  {Abbott}} \emph {et~al.} (\bibinfo {collaboration} {LIGO Scientific, Virgo,
  Fermi-GBM, INTEGRAL}),\ }\href {https://doi.org/10.3847/2041-8213/aa920c}
  {\bibfield  {journal} {\bibinfo  {journal} {Astrophys. J. Lett.}\ }\textbf
  {\bibinfo {volume} {848}},\ \bibinfo {pages} {L13} (\bibinfo {year}
  {2017}{\natexlab{b}})},\ \Eprint {https://arxiv.org/abs/1710.05834}
  {arXiv:1710.05834 [astro-ph.HE]} \BibitemShut {NoStop}%
\bibitem [{\citenamefont {Wang}\ \emph {et~al.}(2017)\citenamefont {Wang} \emph
  {et~al.}}]{Wang:2017rpx}%
  \BibitemOpen
  \bibfield  {author} {\bibinfo {author} {\bibfnamefont {H.}~\bibnamefont
  {Wang}} \emph {et~al.},\ }\href {https://doi.org/10.3847/2041-8213/aa9e08}
  {\bibfield  {journal} {\bibinfo  {journal} {Astrophys. J. Lett.}\ }\textbf
  {\bibinfo {volume} {851}},\ \bibinfo {pages} {L18} (\bibinfo {year}
  {2017})},\ \Eprint {https://arxiv.org/abs/1710.05805} {arXiv:1710.05805
  [astro-ph.HE]} \BibitemShut {NoStop}%
\bibitem [{\citenamefont {Belgacem}\ \emph
  {et~al.}(2018{\natexlab{a}})\citenamefont {Belgacem}, \citenamefont {Dirian},
  \citenamefont {Foffa},\ and\ \citenamefont {Maggiore}}]{Belgacem:2018lbp}%
  \BibitemOpen
  \bibfield  {author} {\bibinfo {author} {\bibfnamefont {E.}~\bibnamefont
  {Belgacem}}, \bibinfo {author} {\bibfnamefont {Y.}~\bibnamefont {Dirian}},
  \bibinfo {author} {\bibfnamefont {S.}~\bibnamefont {Foffa}},\ and\ \bibinfo
  {author} {\bibfnamefont {M.}~\bibnamefont {Maggiore}},\ }\href
  {https://doi.org/10.1103/PhysRevD.98.023510} {\bibfield  {journal} {\bibinfo
  {journal} {Phys. Rev. D}\ }\textbf {\bibinfo {volume} {98}},\ \bibinfo
  {pages} {023510} (\bibinfo {year} {2018}{\natexlab{a}})},\ \Eprint
  {https://arxiv.org/abs/1805.08731} {arXiv:1805.08731 [gr-qc]} \BibitemShut
  {NoStop}%
\bibitem [{\citenamefont {Ferreira}\ \emph {et~al.}(2022)\citenamefont
  {Ferreira}, \citenamefont {Barreiro}, \citenamefont {Mimoso},\ and\
  \citenamefont {Nunes}}]{Ferreira:2022jcd}%
  \BibitemOpen
  \bibfield  {author} {\bibinfo {author} {\bibfnamefont {J.}~\bibnamefont
  {Ferreira}}, \bibinfo {author} {\bibfnamefont {T.}~\bibnamefont {Barreiro}},
  \bibinfo {author} {\bibfnamefont {J.}~\bibnamefont {Mimoso}},\ and\ \bibinfo
  {author} {\bibfnamefont {N.~J.}\ \bibnamefont {Nunes}},\ }\href
  {https://doi.org/10.1103/PhysRevD.105.123531} {\bibfield  {journal} {\bibinfo
   {journal} {Phys. Rev. D}\ }\textbf {\bibinfo {volume} {105}},\ \bibinfo
  {pages} {123531} (\bibinfo {year} {2022})},\ \Eprint
  {https://arxiv.org/abs/2203.13788} {arXiv:2203.13788 [astro-ph.CO]}
  \BibitemShut {NoStop}%
\bibitem [{\citenamefont {Matos}\ \emph {et~al.}(2021)\citenamefont {Matos},
  \citenamefont {Calv\~ao},\ and\ \citenamefont {Waga}}]{Matos:2021qne}%
  \BibitemOpen
  \bibfield  {author} {\bibinfo {author} {\bibfnamefont {I.~S.}\ \bibnamefont
  {Matos}}, \bibinfo {author} {\bibfnamefont {M.~O.}\ \bibnamefont
  {Calv\~ao}},\ and\ \bibinfo {author} {\bibfnamefont {I.}~\bibnamefont
  {Waga}},\ }\href {https://doi.org/10.1103/PhysRevD.103.104059} {\bibfield
  {journal} {\bibinfo  {journal} {Phys. Rev. D}\ }\textbf {\bibinfo {volume}
  {103}},\ \bibinfo {pages} {104059} (\bibinfo {year} {2021})},\ \Eprint
  {https://arxiv.org/abs/2104.10305} {arXiv:2104.10305 [gr-qc]} \BibitemShut
  {NoStop}%
\bibitem [{\citenamefont {Finke}\ \emph {et~al.}(2021)\citenamefont {Finke},
  \citenamefont {Foffa}, \citenamefont {Iacovelli}, \citenamefont {Maggiore},\
  and\ \citenamefont {Mancarella}}]{Finke:2021aom}%
  \BibitemOpen
  \bibfield  {author} {\bibinfo {author} {\bibfnamefont {A.}~\bibnamefont
  {Finke}}, \bibinfo {author} {\bibfnamefont {S.}~\bibnamefont {Foffa}},
  \bibinfo {author} {\bibfnamefont {F.}~\bibnamefont {Iacovelli}}, \bibinfo
  {author} {\bibfnamefont {M.}~\bibnamefont {Maggiore}},\ and\ \bibinfo
  {author} {\bibfnamefont {M.}~\bibnamefont {Mancarella}},\ }\href
  {https://doi.org/10.1088/1475-7516/2021/08/026} {\bibfield  {journal}
  {\bibinfo  {journal} {JCAP}\ }\textbf {\bibinfo {volume} {08}},\ \bibinfo
  {pages} {026}},\ \Eprint {https://arxiv.org/abs/2101.12660} {arXiv:2101.12660
  [astro-ph.CO]} \BibitemShut {NoStop}%
\bibitem [{\citenamefont {Belgacem}\ \emph {et~al.}(2019)\citenamefont
  {Belgacem} \emph {et~al.}}]{LISACosmologyWorkingGroup:2019mwx}%
  \BibitemOpen
  \bibfield  {author} {\bibinfo {author} {\bibfnamefont {E.}~\bibnamefont
  {Belgacem}} \emph {et~al.} (\bibinfo {collaboration} {LISA Cosmology Working
  Group}),\ }\href {https://doi.org/10.1088/1475-7516/2019/07/024} {\bibfield
  {journal} {\bibinfo  {journal} {JCAP}\ }\textbf {\bibinfo {volume} {07}},\
  \bibinfo {pages} {024}},\ \Eprint {https://arxiv.org/abs/1906.01593}
  {arXiv:1906.01593 [astro-ph.CO]} \BibitemShut {NoStop}%
\bibitem [{\citenamefont {Mastrogiovanni}\ \emph {et~al.}(2021)\citenamefont
  {Mastrogiovanni}, \citenamefont {Haegel}, \citenamefont {Karathanasis},
  \citenamefont {Hernandez},\ and\ \citenamefont
  {Steer}}]{Mastrogiovanni:2020mvm}%
  \BibitemOpen
  \bibfield  {author} {\bibinfo {author} {\bibfnamefont {S.}~\bibnamefont
  {Mastrogiovanni}}, \bibinfo {author} {\bibfnamefont {L.}~\bibnamefont
  {Haegel}}, \bibinfo {author} {\bibfnamefont {C.}~\bibnamefont
  {Karathanasis}}, \bibinfo {author} {\bibfnamefont {I.~M.~n.}\ \bibnamefont
  {Hernandez}},\ and\ \bibinfo {author} {\bibfnamefont {D.~A.}\ \bibnamefont
  {Steer}},\ }\href {https://doi.org/10.1088/1475-7516/2021/02/043} {\bibfield
  {journal} {\bibinfo  {journal} {JCAP}\ }\textbf {\bibinfo {volume} {02}},\
  \bibinfo {pages} {043}},\ \Eprint {https://arxiv.org/abs/2010.04047}
  {arXiv:2010.04047 [gr-qc]} \BibitemShut {NoStop}%
\bibitem [{\citenamefont {{Zhu}}\ \emph {et~al.}(2023)\citenamefont {{Zhu}},
  \citenamefont {{Zhao}}, \citenamefont {{Yan}}, \citenamefont {{Gong}},\ and\
  \citenamefont {{Wang}}}]{2023arXiv230409025Z}%
  \BibitemOpen
  \bibfield  {author} {\bibinfo {author} {\bibfnamefont {T.}~\bibnamefont
  {{Zhu}}}, \bibinfo {author} {\bibfnamefont {W.}~\bibnamefont {{Zhao}}},
  \bibinfo {author} {\bibfnamefont {J.-M.}\ \bibnamefont {{Yan}}}, \bibinfo
  {author} {\bibfnamefont {C.}~\bibnamefont {{Gong}}},\ and\ \bibinfo {author}
  {\bibfnamefont {A.}~\bibnamefont {{Wang}}},\ }\href
  {https://doi.org/10.48550/arXiv.2304.09025} {\bibfield  {journal} {\bibinfo
  {journal} {arXiv e-prints}\ ,\ \bibinfo {eid} {arXiv:2304.09025}} (\bibinfo
  {year} {2023})},\ \Eprint {https://arxiv.org/abs/2304.09025}
  {arXiv:2304.09025 [gr-qc]} \BibitemShut {NoStop}%
\bibitem [{\citenamefont {Ren}\ \emph {et~al.}(2022)\citenamefont {Ren},
  \citenamefont {Yan}, \citenamefont {Zhao}, \citenamefont {Cai},\ and\
  \citenamefont {Saridakis}}]{Ren:2022aeo}%
  \BibitemOpen
  \bibfield  {author} {\bibinfo {author} {\bibfnamefont {X.}~\bibnamefont
  {Ren}}, \bibinfo {author} {\bibfnamefont {S.-F.}\ \bibnamefont {Yan}},
  \bibinfo {author} {\bibfnamefont {Y.}~\bibnamefont {Zhao}}, \bibinfo {author}
  {\bibfnamefont {Y.-F.}\ \bibnamefont {Cai}},\ and\ \bibinfo {author}
  {\bibfnamefont {E.~N.}\ \bibnamefont {Saridakis}},\ }\href
  {https://doi.org/10.3847/1538-4357/ac6ba5} {\bibfield  {journal} {\bibinfo
  {journal} {Astrophys. J.}\ }\textbf {\bibinfo {volume} {932}},\ \bibinfo
  {pages} {2} (\bibinfo {year} {2022})},\ \Eprint
  {https://arxiv.org/abs/2203.01926} {arXiv:2203.01926 [astro-ph.CO]}
  \BibitemShut {NoStop}%
\bibitem [{\citenamefont {Wang}\ and\ \citenamefont
  {Mota}(2020)}]{Wang:2020zfv}%
  \BibitemOpen
  \bibfield  {author} {\bibinfo {author} {\bibfnamefont {D.}~\bibnamefont
  {Wang}}\ and\ \bibinfo {author} {\bibfnamefont {D.}~\bibnamefont {Mota}},\
  }\href {https://doi.org/10.1103/PhysRevD.102.063530} {\bibfield  {journal}
  {\bibinfo  {journal} {Phys. Rev. D}\ }\textbf {\bibinfo {volume} {102}},\
  \bibinfo {pages} {063530} (\bibinfo {year} {2020})},\ \Eprint
  {https://arxiv.org/abs/2003.10095} {arXiv:2003.10095 [astro-ph.CO]}
  \BibitemShut {NoStop}%
\bibitem [{\citenamefont {Maggiore}\ \emph {et~al.}(2020)\citenamefont
  {Maggiore} \emph {et~al.}}]{Maggiore:2019uih}%
  \BibitemOpen
  \bibfield  {author} {\bibinfo {author} {\bibfnamefont {M.}~\bibnamefont
  {Maggiore}} \emph {et~al.},\ }\href
  {https://doi.org/10.1088/1475-7516/2020/03/050} {\bibfield  {journal}
  {\bibinfo  {journal} {JCAP}\ }\textbf {\bibinfo {volume} {03}},\ \bibinfo
  {pages} {050}},\ \Eprint {https://arxiv.org/abs/1912.02622} {arXiv:1912.02622
  [astro-ph.CO]} \BibitemShut {NoStop}%
\bibitem [{\citenamefont {Sathyaprakash}\ \emph {et~al.}(2010)\citenamefont
  {Sathyaprakash}, \citenamefont {Schutz},\ and\ \citenamefont {Van
  Den~Broeck}}]{Sathyaprakash:2009xt}%
  \BibitemOpen
  \bibfield  {author} {\bibinfo {author} {\bibfnamefont {B.~S.}\ \bibnamefont
  {Sathyaprakash}}, \bibinfo {author} {\bibfnamefont {B.~F.}\ \bibnamefont
  {Schutz}},\ and\ \bibinfo {author} {\bibfnamefont {C.}~\bibnamefont {Van
  Den~Broeck}},\ }\href {https://doi.org/10.1088/0264-9381/27/21/215006}
  {\bibfield  {journal} {\bibinfo  {journal} {Class. Quant. Grav.}\ }\textbf
  {\bibinfo {volume} {27}},\ \bibinfo {pages} {215006} (\bibinfo {year}
  {2010})},\ \Eprint {https://arxiv.org/abs/0906.4151} {arXiv:0906.4151
  [astro-ph.CO]} \BibitemShut {NoStop}%
\bibitem [{\citenamefont {{Evans}}\ \emph {et~al.}(2021)\citenamefont
  {{Evans}}, \citenamefont {{Adhikari}}, \citenamefont {{Afle}}, \citenamefont
  {{Ballmer}}, \citenamefont {{Biscoveanu}}, \citenamefont {{Borhanian}},
  \citenamefont {{Brown}}, \citenamefont {{Chen}}, \citenamefont
  {{Eisenstein}}, \citenamefont {{Gruson}}, \citenamefont {{Gupta}},
  \citenamefont {{Hall}}, \citenamefont {{Huxford}}, \citenamefont {{Kamai}},
  \citenamefont {{Kashyap}}, \citenamefont {{Kissel}}, \citenamefont {{Kuns}},
  \citenamefont {{Landry}}, \citenamefont {{Lenon}}, \citenamefont
  {{Lovelace}}, \citenamefont {{McCuller}}, \citenamefont {{Ng}}, \citenamefont
  {{Nitz}}, \citenamefont {{Read}}, \citenamefont {{Sathyaprakash}},
  \citenamefont {{Shoemaker}}, \citenamefont {{Slagmolen}}, \citenamefont
  {{Smith}}, \citenamefont {{Srivastava}}, \citenamefont {{Sun}}, \citenamefont
  {{Vitale}},\ and\ \citenamefont {{Weiss}}}]{2021arXiv210909882E}%
  \BibitemOpen
  \bibfield  {author} {\bibinfo {author} {\bibfnamefont {M.}~\bibnamefont
  {{Evans}}}, \bibinfo {author} {\bibfnamefont {R.~X.}\ \bibnamefont
  {{Adhikari}}}, \bibinfo {author} {\bibfnamefont {C.}~\bibnamefont {{Afle}}},
  \bibinfo {author} {\bibfnamefont {S.~W.}\ \bibnamefont {{Ballmer}}}, \bibinfo
  {author} {\bibfnamefont {S.}~\bibnamefont {{Biscoveanu}}}, \bibinfo {author}
  {\bibfnamefont {S.}~\bibnamefont {{Borhanian}}}, \bibinfo {author}
  {\bibfnamefont {D.~A.}\ \bibnamefont {{Brown}}}, \bibinfo {author}
  {\bibfnamefont {Y.}~\bibnamefont {{Chen}}}, \bibinfo {author} {\bibfnamefont
  {R.}~\bibnamefont {{Eisenstein}}}, \bibinfo {author} {\bibfnamefont
  {A.}~\bibnamefont {{Gruson}}}, \bibinfo {author} {\bibfnamefont
  {A.}~\bibnamefont {{Gupta}}}, \bibinfo {author} {\bibfnamefont {E.~D.}\
  \bibnamefont {{Hall}}}, \bibinfo {author} {\bibfnamefont {R.}~\bibnamefont
  {{Huxford}}}, \bibinfo {author} {\bibfnamefont {B.}~\bibnamefont {{Kamai}}},
  \bibinfo {author} {\bibfnamefont {R.}~\bibnamefont {{Kashyap}}}, \bibinfo
  {author} {\bibfnamefont {J.~S.}\ \bibnamefont {{Kissel}}}, \bibinfo {author}
  {\bibfnamefont {K.}~\bibnamefont {{Kuns}}}, \bibinfo {author} {\bibfnamefont
  {P.}~\bibnamefont {{Landry}}}, \bibinfo {author} {\bibfnamefont
  {A.}~\bibnamefont {{Lenon}}}, \bibinfo {author} {\bibfnamefont
  {G.}~\bibnamefont {{Lovelace}}}, \bibinfo {author} {\bibfnamefont
  {L.}~\bibnamefont {{McCuller}}}, \bibinfo {author} {\bibfnamefont {K.~K.~Y.}\
  \bibnamefont {{Ng}}}, \bibinfo {author} {\bibfnamefont {A.~H.}\ \bibnamefont
  {{Nitz}}}, \bibinfo {author} {\bibfnamefont {J.}~\bibnamefont {{Read}}},
  \bibinfo {author} {\bibfnamefont {B.~S.}\ \bibnamefont {{Sathyaprakash}}},
  \bibinfo {author} {\bibfnamefont {D.~H.}\ \bibnamefont {{Shoemaker}}},
  \bibinfo {author} {\bibfnamefont {B.~J.~J.}\ \bibnamefont {{Slagmolen}}},
  \bibinfo {author} {\bibfnamefont {J.~R.}\ \bibnamefont {{Smith}}}, \bibinfo
  {author} {\bibfnamefont {V.}~\bibnamefont {{Srivastava}}}, \bibinfo {author}
  {\bibfnamefont {L.}~\bibnamefont {{Sun}}}, \bibinfo {author} {\bibfnamefont
  {S.}~\bibnamefont {{Vitale}}},\ and\ \bibinfo {author} {\bibfnamefont
  {R.}~\bibnamefont {{Weiss}}},\ }\href
  {https://doi.org/10.48550/arXiv.2109.09882} {\bibfield  {journal} {\bibinfo
  {journal} {arXiv e-prints}\ ,\ \bibinfo {eid} {arXiv:2109.09882}} (\bibinfo
  {year} {2021})},\ \Eprint {https://arxiv.org/abs/2109.09882}
  {arXiv:2109.09882 [astro-ph.IM]} \BibitemShut {NoStop}%
\bibitem [{\citenamefont {{Wang}}\ \emph {et~al.}(2023)\citenamefont {{Wang}},
  \citenamefont {{Tang}}, \citenamefont {{Jin}},\ and\ \citenamefont
  {{Fan}}}]{2023ApJ...943...13W}%
  \BibitemOpen
  \bibfield  {author} {\bibinfo {author} {\bibfnamefont {Y.-Y.}\ \bibnamefont
  {{Wang}}}, \bibinfo {author} {\bibfnamefont {S.-P.}\ \bibnamefont {{Tang}}},
  \bibinfo {author} {\bibfnamefont {Z.-P.}\ \bibnamefont {{Jin}}},\ and\
  \bibinfo {author} {\bibfnamefont {Y.-Z.}\ \bibnamefont {{Fan}}},\ }\href
  {https://doi.org/10.3847/1538-4357/aca96c} {\bibfield  {journal} {\bibinfo
  {journal} {\apj}\ }\textbf {\bibinfo {volume} {943}},\ \bibinfo {eid} {13}
  (\bibinfo {year} {2023})},\ \Eprint {https://arxiv.org/abs/2208.09121}
  {arXiv:2208.09121 [astro-ph.HE]} \BibitemShut {NoStop}%
\bibitem [{\citenamefont {Bengochea}\ and\ \citenamefont
  {Ferraro}(2009)}]{Bengochea:2008gz}%
  \BibitemOpen
  \bibfield  {author} {\bibinfo {author} {\bibfnamefont {G.~R.}\ \bibnamefont
  {Bengochea}}\ and\ \bibinfo {author} {\bibfnamefont {R.}~\bibnamefont
  {Ferraro}},\ }\href {https://doi.org/10.1103/PhysRevD.79.124019} {\bibfield
  {journal} {\bibinfo  {journal} {Phys. Rev. D}\ }\textbf {\bibinfo {volume}
  {79}},\ \bibinfo {pages} {124019} (\bibinfo {year} {2009})},\ \Eprint
  {https://arxiv.org/abs/0812.1205} {arXiv:0812.1205 [astro-ph]} \BibitemShut
  {NoStop}%
\bibitem [{\citenamefont {Li}\ \emph {et~al.}(2018)\citenamefont {Li},
  \citenamefont {Cai}, \citenamefont {Cai},\ and\ \citenamefont
  {Saridakis}}]{Li:2018ixg}%
  \BibitemOpen
  \bibfield  {author} {\bibinfo {author} {\bibfnamefont {C.}~\bibnamefont
  {Li}}, \bibinfo {author} {\bibfnamefont {Y.}~\bibnamefont {Cai}}, \bibinfo
  {author} {\bibfnamefont {Y.-F.}\ \bibnamefont {Cai}},\ and\ \bibinfo {author}
  {\bibfnamefont {E.~N.}\ \bibnamefont {Saridakis}},\ }\href
  {https://doi.org/10.1088/1475-7516/2018/10/001} {\bibfield  {journal}
  {\bibinfo  {journal} {JCAP}\ }\textbf {\bibinfo {volume} {10}},\ \bibinfo
  {pages} {001}},\ \Eprint {https://arxiv.org/abs/1803.09818} {arXiv:1803.09818
  [gr-qc]} \BibitemShut {NoStop}%
\bibitem [{\citenamefont {Chen}\ \emph {et~al.}(2011)\citenamefont {Chen},
  \citenamefont {Dent}, \citenamefont {Dutta},\ and\ \citenamefont
  {Saridakis}}]{Chen:2010va}%
  \BibitemOpen
  \bibfield  {author} {\bibinfo {author} {\bibfnamefont {S.-H.}\ \bibnamefont
  {Chen}}, \bibinfo {author} {\bibfnamefont {J.~B.}\ \bibnamefont {Dent}},
  \bibinfo {author} {\bibfnamefont {S.}~\bibnamefont {Dutta}},\ and\ \bibinfo
  {author} {\bibfnamefont {E.~N.}\ \bibnamefont {Saridakis}},\ }\href
  {https://doi.org/10.1103/PhysRevD.83.023508} {\bibfield  {journal} {\bibinfo
  {journal} {Phys. Rev. D}\ }\textbf {\bibinfo {volume} {83}},\ \bibinfo
  {pages} {023508} (\bibinfo {year} {2011})},\ \Eprint
  {https://arxiv.org/abs/1008.1250} {arXiv:1008.1250 [astro-ph.CO]}
  \BibitemShut {NoStop}%
\bibitem [{\citenamefont {Farrugia}\ \emph {et~al.}(2018)\citenamefont
  {Farrugia}, \citenamefont {Levi~Said}, \citenamefont {Gakis},\ and\
  \citenamefont {Saridakis}}]{Farrugia:2018gyz}%
  \BibitemOpen
  \bibfield  {author} {\bibinfo {author} {\bibfnamefont {G.}~\bibnamefont
  {Farrugia}}, \bibinfo {author} {\bibfnamefont {J.}~\bibnamefont {Levi~Said}},
  \bibinfo {author} {\bibfnamefont {V.}~\bibnamefont {Gakis}},\ and\ \bibinfo
  {author} {\bibfnamefont {E.~N.}\ \bibnamefont {Saridakis}},\ }\href
  {https://doi.org/10.1103/PhysRevD.97.124064} {\bibfield  {journal} {\bibinfo
  {journal} {Phys. Rev. D}\ }\textbf {\bibinfo {volume} {97}},\ \bibinfo
  {pages} {124064} (\bibinfo {year} {2018})},\ \Eprint
  {https://arxiv.org/abs/1804.07365} {arXiv:1804.07365 [gr-qc]} \BibitemShut
  {NoStop}%
\bibitem [{\citenamefont {Belgacem}\ \emph
  {et~al.}(2018{\natexlab{b}})\citenamefont {Belgacem}, \citenamefont {Dirian},
  \citenamefont {Foffa},\ and\ \citenamefont {Maggiore}}]{Belgacem:2017ihm}%
  \BibitemOpen
  \bibfield  {author} {\bibinfo {author} {\bibfnamefont {E.}~\bibnamefont
  {Belgacem}}, \bibinfo {author} {\bibfnamefont {Y.}~\bibnamefont {Dirian}},
  \bibinfo {author} {\bibfnamefont {S.}~\bibnamefont {Foffa}},\ and\ \bibinfo
  {author} {\bibfnamefont {M.}~\bibnamefont {Maggiore}},\ }\href
  {https://doi.org/10.1103/PhysRevD.97.104066} {\bibfield  {journal} {\bibinfo
  {journal} {Phys. Rev. D}\ }\textbf {\bibinfo {volume} {97}},\ \bibinfo
  {pages} {104066} (\bibinfo {year} {2018}{\natexlab{b}})},\ \Eprint
  {https://arxiv.org/abs/1712.08108} {arXiv:1712.08108 [astro-ph.CO]}
  \BibitemShut {NoStop}%
\bibitem [{\citenamefont {Branch}\ and\ \citenamefont
  {Tammann}(1992)}]{Branch:1992rv}%
  \BibitemOpen
  \bibfield  {author} {\bibinfo {author} {\bibfnamefont {D.}~\bibnamefont
  {Branch}}\ and\ \bibinfo {author} {\bibfnamefont {G.~A.}\ \bibnamefont
  {Tammann}},\ }\href {https://doi.org/10.1146/annurev.aa.30.090192.002043}
  {\bibfield  {journal} {\bibinfo  {journal} {Ann. Rev. Astron. Astrophys.}\
  }\textbf {\bibinfo {volume} {30}},\ \bibinfo {pages} {359} (\bibinfo {year}
  {1992})}\BibitemShut {NoStop}%
\bibitem [{\citenamefont {Wright}\ and\ \citenamefont
  {Li}(2018)}]{Wright:2017rsu}%
  \BibitemOpen
  \bibfield  {author} {\bibinfo {author} {\bibfnamefont {B.~S.}\ \bibnamefont
  {Wright}}\ and\ \bibinfo {author} {\bibfnamefont {B.}~\bibnamefont {Li}},\
  }\href {https://doi.org/10.1103/PhysRevD.97.083505} {\bibfield  {journal}
  {\bibinfo  {journal} {Phys. Rev. D}\ }\textbf {\bibinfo {volume} {97}},\
  \bibinfo {pages} {083505} (\bibinfo {year} {2018})},\ \Eprint
  {https://arxiv.org/abs/1710.07018} {arXiv:1710.07018 [astro-ph.CO]}
  \BibitemShut {NoStop}%
\bibitem [{\citenamefont {Scolnic}\ \emph {et~al.}(2018)\citenamefont {Scolnic}
  \emph {et~al.}}]{Pan-STARRS1:2017jku}%
  \BibitemOpen
  \bibfield  {author} {\bibinfo {author} {\bibfnamefont {D.~M.}\ \bibnamefont
  {Scolnic}} \emph {et~al.} (\bibinfo {collaboration} {Pan-STARRS1}),\ }\href
  {https://doi.org/10.3847/1538-4357/aab9bb} {\bibfield  {journal} {\bibinfo
  {journal} {Astrophys. J.}\ }\textbf {\bibinfo {volume} {859}},\ \bibinfo
  {pages} {101} (\bibinfo {year} {2018})},\ \Eprint
  {https://arxiv.org/abs/1710.00845} {arXiv:1710.00845 [astro-ph.CO]}
  \BibitemShut {NoStop}%
\bibitem [{\citenamefont {Zhu}\ \emph {et~al.}(2021)\citenamefont {Zhu} \emph
  {et~al.}}]{Zhu:2020ffa}%
  \BibitemOpen
  \bibfield  {author} {\bibinfo {author} {\bibfnamefont {J.-P.}\ \bibnamefont
  {Zhu}} \emph {et~al.},\ }\href {https://doi.org/10.3847/1538-4357/abfe5e}
  {\bibfield  {journal} {\bibinfo  {journal} {Astrophys. J.}\ }\textbf
  {\bibinfo {volume} {917}},\ \bibinfo {pages} {24} (\bibinfo {year} {2021})},\
  \Eprint {https://arxiv.org/abs/2011.02717} {arXiv:2011.02717 [astro-ph.HE]}
  \BibitemShut {NoStop}%
\bibitem [{\citenamefont {Abbott}\ \emph {et~al.}(2019)\citenamefont {Abbott}
  \emph {et~al.}}]{LIGOScientific:2018hze}%
  \BibitemOpen
  \bibfield  {author} {\bibinfo {author} {\bibfnamefont {B.~P.}\ \bibnamefont
  {Abbott}} \emph {et~al.} (\bibinfo {collaboration} {LIGO Scientific,
  Virgo}),\ }\href {https://doi.org/10.1103/PhysRevX.9.011001} {\bibfield
  {journal} {\bibinfo  {journal} {Phys. Rev. X}\ }\textbf {\bibinfo {volume}
  {9}},\ \bibinfo {pages} {011001} (\bibinfo {year} {2019})},\ \Eprint
  {https://arxiv.org/abs/1805.11579} {arXiv:1805.11579 [gr-qc]} \BibitemShut
  {NoStop}%
\bibitem [{\citenamefont {Dietrich}\ \emph {et~al.}(2019)\citenamefont
  {Dietrich} \emph {et~al.}}]{Dietrich:2018uni}%
  \BibitemOpen
  \bibfield  {author} {\bibinfo {author} {\bibfnamefont {T.}~\bibnamefont
  {Dietrich}} \emph {et~al.},\ }\href
  {https://doi.org/10.1103/PhysRevD.99.024029} {\bibfield  {journal} {\bibinfo
  {journal} {Phys. Rev. D}\ }\textbf {\bibinfo {volume} {99}},\ \bibinfo
  {pages} {024029} (\bibinfo {year} {2019})},\ \Eprint
  {https://arxiv.org/abs/1804.02235} {arXiv:1804.02235 [gr-qc]} \BibitemShut
  {NoStop}%
\bibitem [{\citenamefont {Thompson}\ \emph {et~al.}(2020)\citenamefont
  {Thompson}, \citenamefont {Fauchon-Jones}, \citenamefont {Khan},
  \citenamefont {Nitoglia}, \citenamefont {Pannarale}, \citenamefont
  {Dietrich},\ and\ \citenamefont {Hannam}}]{Thompson:2020nei}%
  \BibitemOpen
  \bibfield  {author} {\bibinfo {author} {\bibfnamefont {J.~E.}\ \bibnamefont
  {Thompson}}, \bibinfo {author} {\bibfnamefont {E.}~\bibnamefont
  {Fauchon-Jones}}, \bibinfo {author} {\bibfnamefont {S.}~\bibnamefont {Khan}},
  \bibinfo {author} {\bibfnamefont {E.}~\bibnamefont {Nitoglia}}, \bibinfo
  {author} {\bibfnamefont {F.}~\bibnamefont {Pannarale}}, \bibinfo {author}
  {\bibfnamefont {T.}~\bibnamefont {Dietrich}},\ and\ \bibinfo {author}
  {\bibfnamefont {M.}~\bibnamefont {Hannam}},\ }\href
  {https://doi.org/10.1103/PhysRevD.101.124059} {\bibfield  {journal} {\bibinfo
   {journal} {Phys. Rev. D}\ }\textbf {\bibinfo {volume} {101}},\ \bibinfo
  {pages} {124059} (\bibinfo {year} {2020})},\ \Eprint
  {https://arxiv.org/abs/2002.08383} {arXiv:2002.08383 [gr-qc]} \BibitemShut
  {NoStop}%
\bibitem [{\citenamefont {Iacovelli}\ \emph {et~al.}(2022)\citenamefont
  {Iacovelli}, \citenamefont {Mancarella}, \citenamefont {Foffa},\ and\
  \citenamefont {Maggiore}}]{Iacovelli:2022bbs}%
  \BibitemOpen
  \bibfield  {author} {\bibinfo {author} {\bibfnamefont {F.}~\bibnamefont
  {Iacovelli}}, \bibinfo {author} {\bibfnamefont {M.}~\bibnamefont
  {Mancarella}}, \bibinfo {author} {\bibfnamefont {S.}~\bibnamefont {Foffa}},\
  and\ \bibinfo {author} {\bibfnamefont {M.}~\bibnamefont {Maggiore}},\ }\href
  {https://doi.org/10.3847/1538-4357/ac9cd4} {\bibfield  {journal} {\bibinfo
  {journal} {Astrophys. J.}\ }\textbf {\bibinfo {volume} {941}},\ \bibinfo
  {pages} {208} (\bibinfo {year} {2022})},\ \Eprint
  {https://arxiv.org/abs/2207.02771} {arXiv:2207.02771 [gr-qc]} \BibitemShut
  {NoStop}%
\bibitem [{\citenamefont {Dawson}\ \emph {et~al.}(2013)\citenamefont {Dawson}
  \emph {et~al.}}]{BOSS:2012dmf}%
  \BibitemOpen
  \bibfield  {author} {\bibinfo {author} {\bibfnamefont {K.~S.}\ \bibnamefont
  {Dawson}} \emph {et~al.} (\bibinfo {collaboration} {BOSS}),\ }\href
  {https://doi.org/10.1088/0004-6256/145/1/10} {\bibfield  {journal} {\bibinfo
  {journal} {Astron. J.}\ }\textbf {\bibinfo {volume} {145}},\ \bibinfo {pages}
  {10} (\bibinfo {year} {2013})},\ \Eprint {https://arxiv.org/abs/1208.0022}
  {arXiv:1208.0022 [astro-ph.CO]} \BibitemShut {NoStop}%
\bibitem [{\citenamefont {Dawson}\ \emph {et~al.}(2016)\citenamefont {Dawson}
  \emph {et~al.}}]{Dawson:2015wdb}%
  \BibitemOpen
  \bibfield  {author} {\bibinfo {author} {\bibfnamefont {K.~S.}\ \bibnamefont
  {Dawson}} \emph {et~al.},\ }\href
  {https://doi.org/10.3847/0004-6256/151/2/44} {\bibfield  {journal} {\bibinfo
  {journal} {Astron. J.}\ }\textbf {\bibinfo {volume} {151}},\ \bibinfo {pages}
  {44} (\bibinfo {year} {2016})},\ \Eprint {https://arxiv.org/abs/1508.04473}
  {arXiv:1508.04473 [astro-ph.CO]} \BibitemShut {NoStop}%
\bibitem [{\citenamefont {Sanchez}\ \emph {et~al.}(2011)\citenamefont
  {Sanchez}, \citenamefont {Carnero}, \citenamefont {Garcia-Bellido},
  \citenamefont {Gaztanaga}, \citenamefont {de~Simoni}, \citenamefont {Crocce},
  \citenamefont {Cabre}, \citenamefont {Fosalba},\ and\ \citenamefont
  {Alonso}}]{Sanchez:2010zg}%
  \BibitemOpen
  \bibfield  {author} {\bibinfo {author} {\bibfnamefont {E.}~\bibnamefont
  {Sanchez}}, \bibinfo {author} {\bibfnamefont {A.}~\bibnamefont {Carnero}},
  \bibinfo {author} {\bibfnamefont {J.}~\bibnamefont {Garcia-Bellido}},
  \bibinfo {author} {\bibfnamefont {E.}~\bibnamefont {Gaztanaga}}, \bibinfo
  {author} {\bibfnamefont {F.}~\bibnamefont {de~Simoni}}, \bibinfo {author}
  {\bibfnamefont {M.}~\bibnamefont {Crocce}}, \bibinfo {author} {\bibfnamefont
  {A.}~\bibnamefont {Cabre}}, \bibinfo {author} {\bibfnamefont
  {P.}~\bibnamefont {Fosalba}},\ and\ \bibinfo {author} {\bibfnamefont
  {D.}~\bibnamefont {Alonso}},\ }\href
  {https://doi.org/10.1111/j.1365-2966.2010.17679.x} {\bibfield  {journal}
  {\bibinfo  {journal} {Mon. Not. Roy. Astron. Soc.}\ }\textbf {\bibinfo
  {volume} {411}},\ \bibinfo {pages} {277} (\bibinfo {year} {2011})},\ \Eprint
  {https://arxiv.org/abs/1006.3226} {arXiv:1006.3226 [astro-ph.CO]}
  \BibitemShut {NoStop}%
\bibitem [{\citenamefont {Carvalho}\ \emph {et~al.}(2020)\citenamefont
  {Carvalho}, \citenamefont {Bernui}, \citenamefont {Benetti}, \citenamefont
  {Carvalho}, \citenamefont {de~Carvalho},\ and\ \citenamefont
  {Alcaniz}}]{Carvalho:2017tuu}%
  \BibitemOpen
  \bibfield  {author} {\bibinfo {author} {\bibfnamefont {G.~C.}\ \bibnamefont
  {Carvalho}}, \bibinfo {author} {\bibfnamefont {A.}~\bibnamefont {Bernui}},
  \bibinfo {author} {\bibfnamefont {M.}~\bibnamefont {Benetti}}, \bibinfo
  {author} {\bibfnamefont {J.~C.}\ \bibnamefont {Carvalho}}, \bibinfo {author}
  {\bibfnamefont {E.}~\bibnamefont {de~Carvalho}},\ and\ \bibinfo {author}
  {\bibfnamefont {J.~S.}\ \bibnamefont {Alcaniz}},\ }\href
  {https://doi.org/10.1016/j.astropartphys.2020.102432} {\bibfield  {journal}
  {\bibinfo  {journal} {Astropart. Phys.}\ }\textbf {\bibinfo {volume} {119}},\
  \bibinfo {pages} {102432} (\bibinfo {year} {2020})},\ \Eprint
  {https://arxiv.org/abs/1709.00271} {arXiv:1709.00271 [astro-ph.CO]}
  \BibitemShut {NoStop}%
\bibitem [{\citenamefont {Menote}\ and\ \citenamefont
  {Marra}(2022)}]{Menote:2021jaq}%
  \BibitemOpen
  \bibfield  {author} {\bibinfo {author} {\bibfnamefont {R.}~\bibnamefont
  {Menote}}\ and\ \bibinfo {author} {\bibfnamefont {V.}~\bibnamefont {Marra}},\
  }\href {https://doi.org/10.1093/mnras/stac847} {\bibfield  {journal}
  {\bibinfo  {journal} {Mon. Not. Roy. Astron. Soc.}\ }\textbf {\bibinfo
  {volume} {513}},\ \bibinfo {pages} {1600} (\bibinfo {year} {2022})},\ \Eprint
  {https://arxiv.org/abs/2112.10000} {arXiv:2112.10000 [astro-ph.CO]}
  \BibitemShut {NoStop}%
\bibitem [{\citenamefont {{Jin}}\ \emph {et~al.}(2018)\citenamefont {{Jin}},
  \citenamefont {{Li}}, \citenamefont {{Wang}}, \citenamefont {{Wang}},
  \citenamefont {{He}}, \citenamefont {{Yuan}}, \citenamefont {{Zhang}},
  \citenamefont {{Zou}}, \citenamefont {{Fan}},\ and\ \citenamefont
  {{Wei}}}]{2018ApJ...857..128J}%
  \BibitemOpen
  \bibfield  {author} {\bibinfo {author} {\bibfnamefont {Z.-P.}\ \bibnamefont
  {{Jin}}}, \bibinfo {author} {\bibfnamefont {X.}~\bibnamefont {{Li}}},
  \bibinfo {author} {\bibfnamefont {H.}~\bibnamefont {{Wang}}}, \bibinfo
  {author} {\bibfnamefont {Y.-Z.}\ \bibnamefont {{Wang}}}, \bibinfo {author}
  {\bibfnamefont {H.-N.}\ \bibnamefont {{He}}}, \bibinfo {author}
  {\bibfnamefont {Q.}~\bibnamefont {{Yuan}}}, \bibinfo {author} {\bibfnamefont
  {F.-W.}\ \bibnamefont {{Zhang}}}, \bibinfo {author} {\bibfnamefont {Y.-C.}\
  \bibnamefont {{Zou}}}, \bibinfo {author} {\bibfnamefont {Y.-Z.}\ \bibnamefont
  {{Fan}}},\ and\ \bibinfo {author} {\bibfnamefont {D.-M.}\ \bibnamefont
  {{Wei}}},\ }\href {https://doi.org/10.3847/1538-4357/aab76d} {\bibfield
  {journal} {\bibinfo  {journal} {\apj}\ }\textbf {\bibinfo {volume} {857}},\
  \bibinfo {eid} {128} (\bibinfo {year} {2018})},\ \Eprint
  {https://arxiv.org/abs/1708.07008} {arXiv:1708.07008 [astro-ph.HE]}
  \BibitemShut {NoStop}%
\bibitem [{\citenamefont {Tang}\ \emph {et~al.}(2023)\citenamefont {Tang},
  \citenamefont {Fan},\ and\ \citenamefont {Wei}}]{Tang:2022qim}%
  \BibitemOpen
  \bibfield  {author} {\bibinfo {author} {\bibfnamefont {S.-P.}\ \bibnamefont
  {Tang}}, \bibinfo {author} {\bibfnamefont {Y.-Z.}\ \bibnamefont {Fan}},\ and\
  \bibinfo {author} {\bibfnamefont {D.-M.}\ \bibnamefont {Wei}},\ }\href
  {https://doi.org/10.1093/mnras/stad1676} {\bibfield  {journal} {\bibinfo
  {journal} {Mon. Not. Roy. Astron. Soc.}\ }\textbf {\bibinfo {volume} {523}},\
  \bibinfo {pages} {4113} (\bibinfo {year} {2023})},\ \Eprint
  {https://arxiv.org/abs/2209.03631} {arXiv:2209.03631 [gr-qc]} \BibitemShut
  {NoStop}%
\bibitem [{\citenamefont {Ozel}\ \emph {et~al.}(2012)\citenamefont {Ozel},
  \citenamefont {Psaltis}, \citenamefont {Narayan},\ and\ \citenamefont
  {Villarreal}}]{Ozel:2012ax}%
  \BibitemOpen
  \bibfield  {author} {\bibinfo {author} {\bibfnamefont {F.}~\bibnamefont
  {Ozel}}, \bibinfo {author} {\bibfnamefont {D.}~\bibnamefont {Psaltis}},
  \bibinfo {author} {\bibfnamefont {R.}~\bibnamefont {Narayan}},\ and\ \bibinfo
  {author} {\bibfnamefont {A.~S.}\ \bibnamefont {Villarreal}},\ }\href
  {https://doi.org/10.1088/0004-637X/757/1/55} {\bibfield  {journal} {\bibinfo
  {journal} {Astrophys. J.}\ }\textbf {\bibinfo {volume} {757}},\ \bibinfo
  {pages} {55} (\bibinfo {year} {2012})},\ \Eprint
  {https://arxiv.org/abs/1201.1006} {arXiv:1201.1006 [astro-ph.HE]}
  \BibitemShut {NoStop}%
\bibitem [{\citenamefont {Kiziltan}\ \emph {et~al.}(2013)\citenamefont
  {Kiziltan}, \citenamefont {Kottas}, \citenamefont {De~Yoreo},\ and\
  \citenamefont {Thorsett}}]{Kiziltan:2013oja}%
  \BibitemOpen
  \bibfield  {author} {\bibinfo {author} {\bibfnamefont {B.}~\bibnamefont
  {Kiziltan}}, \bibinfo {author} {\bibfnamefont {A.}~\bibnamefont {Kottas}},
  \bibinfo {author} {\bibfnamefont {M.}~\bibnamefont {De~Yoreo}},\ and\
  \bibinfo {author} {\bibfnamefont {S.~E.}\ \bibnamefont {Thorsett}},\ }\href
  {https://doi.org/10.1088/0004-637X/778/1/66} {\bibfield  {journal} {\bibinfo
  {journal} {Astrophys. J.}\ }\textbf {\bibinfo {volume} {778}},\ \bibinfo
  {pages} {66} (\bibinfo {year} {2013})},\ \Eprint
  {https://arxiv.org/abs/1309.6635} {arXiv:1309.6635 [astro-ph.SR]}
  \BibitemShut {NoStop}%
\bibitem [{\citenamefont {Shao}\ \emph {et~al.}(2020)\citenamefont {Shao},
  \citenamefont {Tang}, \citenamefont {Jiang},\ and\ \citenamefont
  {Fan}}]{Shao:2020bzt}%
  \BibitemOpen
  \bibfield  {author} {\bibinfo {author} {\bibfnamefont {D.-S.}\ \bibnamefont
  {Shao}}, \bibinfo {author} {\bibfnamefont {S.-P.}\ \bibnamefont {Tang}},
  \bibinfo {author} {\bibfnamefont {J.-L.}\ \bibnamefont {Jiang}},\ and\
  \bibinfo {author} {\bibfnamefont {Y.-Z.}\ \bibnamefont {Fan}},\ }\href
  {https://doi.org/10.1103/PhysRevD.102.063006} {\bibfield  {journal} {\bibinfo
   {journal} {Phys. Rev. D}\ }\textbf {\bibinfo {volume} {102}},\ \bibinfo
  {pages} {063006} (\bibinfo {year} {2020})},\ \Eprint
  {https://arxiv.org/abs/2009.04275} {arXiv:2009.04275 [astro-ph.HE]}
  \BibitemShut {NoStop}%
\bibitem [{\citenamefont {Akmal}\ \emph {et~al.}(1998)\citenamefont {Akmal},
  \citenamefont {Pandharipande},\ and\ \citenamefont
  {Ravenhall}}]{Akmal:1998cf}%
  \BibitemOpen
  \bibfield  {author} {\bibinfo {author} {\bibfnamefont {A.}~\bibnamefont
  {Akmal}}, \bibinfo {author} {\bibfnamefont {V.~R.}\ \bibnamefont
  {Pandharipande}},\ and\ \bibinfo {author} {\bibfnamefont {D.~G.}\
  \bibnamefont {Ravenhall}},\ }\href {https://doi.org/10.1103/PhysRevC.58.1804}
  {\bibfield  {journal} {\bibinfo  {journal} {Phys. Rev. C}\ }\textbf {\bibinfo
  {volume} {58}},\ \bibinfo {pages} {1804} (\bibinfo {year} {1998})},\ \Eprint
  {https://arxiv.org/abs/nucl-th/9804027} {arXiv:nucl-th/9804027} \BibitemShut
  {NoStop}%
\bibitem [{\citenamefont {Foucart}\ \emph {et~al.}(2018)\citenamefont
  {Foucart}, \citenamefont {Hinderer},\ and\ \citenamefont
  {Nissanke}}]{Foucart:2018rjc}%
  \BibitemOpen
  \bibfield  {author} {\bibinfo {author} {\bibfnamefont {F.}~\bibnamefont
  {Foucart}}, \bibinfo {author} {\bibfnamefont {T.}~\bibnamefont {Hinderer}},\
  and\ \bibinfo {author} {\bibfnamefont {S.}~\bibnamefont {Nissanke}},\ }\href
  {https://doi.org/10.1103/PhysRevD.98.081501} {\bibfield  {journal} {\bibinfo
  {journal} {Phys. Rev. D}\ }\textbf {\bibinfo {volume} {98}},\ \bibinfo
  {pages} {081501} (\bibinfo {year} {2018})},\ \Eprint
  {https://arxiv.org/abs/1807.00011} {arXiv:1807.00011 [astro-ph.HE]}
  \BibitemShut {NoStop}%
\bibitem [{\citenamefont {Madau}\ and\ \citenamefont
  {Dickinson}(2014)}]{Madau:2014bja}%
  \BibitemOpen
  \bibfield  {author} {\bibinfo {author} {\bibfnamefont {P.}~\bibnamefont
  {Madau}}\ and\ \bibinfo {author} {\bibfnamefont {M.}~\bibnamefont
  {Dickinson}},\ }\href {https://doi.org/10.1146/annurev-astro-081811-125615}
  {\bibfield  {journal} {\bibinfo  {journal} {Ann. Rev. Astron. Astrophys.}\
  }\textbf {\bibinfo {volume} {52}},\ \bibinfo {pages} {415} (\bibinfo {year}
  {2014})},\ \Eprint {https://arxiv.org/abs/1403.0007} {arXiv:1403.0007
  [astro-ph.CO]} \BibitemShut {NoStop}%
\bibitem [{\citenamefont {Madau}\ and\ \citenamefont
  {Fragos}(2017)}]{Madau:2016jbv}%
  \BibitemOpen
  \bibfield  {author} {\bibinfo {author} {\bibfnamefont {P.}~\bibnamefont
  {Madau}}\ and\ \bibinfo {author} {\bibfnamefont {T.}~\bibnamefont {Fragos}},\
  }\href {https://doi.org/10.3847/1538-4357/aa6af9} {\bibfield  {journal}
  {\bibinfo  {journal} {Astrophys. J.}\ }\textbf {\bibinfo {volume} {840}},\
  \bibinfo {pages} {39} (\bibinfo {year} {2017})},\ \Eprint
  {https://arxiv.org/abs/1606.07887} {arXiv:1606.07887 [astro-ph.GA]}
  \BibitemShut {NoStop}%
\bibitem [{\citenamefont {Mapelli}\ and\ \citenamefont
  {Giacobbo}(2018)}]{Mapelli:2018wys}%
  \BibitemOpen
  \bibfield  {author} {\bibinfo {author} {\bibfnamefont {M.}~\bibnamefont
  {Mapelli}}\ and\ \bibinfo {author} {\bibfnamefont {N.}~\bibnamefont
  {Giacobbo}},\ }\href {https://doi.org/10.1093/mnras/sty1613} {\bibfield
  {journal} {\bibinfo  {journal} {Mon. Not. Roy. Astron. Soc.}\ }\textbf
  {\bibinfo {volume} {479}},\ \bibinfo {pages} {4391} (\bibinfo {year}
  {2018})},\ \Eprint {https://arxiv.org/abs/1806.04866} {arXiv:1806.04866
  [astro-ph.HE]} \BibitemShut {NoStop}%
\bibitem [{\citenamefont {Fryer}\ \emph {et~al.}(2012)\citenamefont {Fryer},
  \citenamefont {Belczynski}, \citenamefont {Wiktorowicz}, \citenamefont
  {Dominik}, \citenamefont {Kalogera},\ and\ \citenamefont
  {Holz}}]{Fryer_2012}%
  \BibitemOpen
  \bibfield  {author} {\bibinfo {author} {\bibfnamefont {C.~L.}\ \bibnamefont
  {Fryer}}, \bibinfo {author} {\bibfnamefont {K.}~\bibnamefont {Belczynski}},
  \bibinfo {author} {\bibfnamefont {G.}~\bibnamefont {Wiktorowicz}}, \bibinfo
  {author} {\bibfnamefont {M.}~\bibnamefont {Dominik}}, \bibinfo {author}
  {\bibfnamefont {V.}~\bibnamefont {Kalogera}},\ and\ \bibinfo {author}
  {\bibfnamefont {D.~E.}\ \bibnamefont {Holz}},\ }\href
  {https://doi.org/10.1088/0004-637x/749/1/91} {\bibfield  {journal} {\bibinfo
  {journal} {The Astrophysical Journal}\ }\textbf {\bibinfo {volume} {749}},\
  \bibinfo {pages} {91} (\bibinfo {year} {2012})}\BibitemShut {NoStop}%
\bibitem [{\citenamefont {Vitale}\ \emph {et~al.}(2017)\citenamefont {Vitale},
  \citenamefont {Lynch}, \citenamefont {Sturani},\ and\ \citenamefont
  {Graff}}]{Vitale:2015tea}%
  \BibitemOpen
  \bibfield  {author} {\bibinfo {author} {\bibfnamefont {S.}~\bibnamefont
  {Vitale}}, \bibinfo {author} {\bibfnamefont {R.}~\bibnamefont {Lynch}},
  \bibinfo {author} {\bibfnamefont {R.}~\bibnamefont {Sturani}},\ and\ \bibinfo
  {author} {\bibfnamefont {P.}~\bibnamefont {Graff}},\ }\href
  {https://doi.org/10.1088/1361-6382/aa552e} {\bibfield  {journal} {\bibinfo
  {journal} {Class. Quant. Grav.}\ }\textbf {\bibinfo {volume} {34}},\ \bibinfo
  {pages} {03LT01} (\bibinfo {year} {2017})},\ \Eprint
  {https://arxiv.org/abs/1503.04307} {arXiv:1503.04307 [gr-qc]} \BibitemShut
  {NoStop}%
\bibitem [{\citenamefont {Lorimer}(2008)}]{Lorimer:2008se}%
  \BibitemOpen
  \bibfield  {author} {\bibinfo {author} {\bibfnamefont {D.~R.}\ \bibnamefont
  {Lorimer}},\ }\href {https://doi.org/10.12942/lrr-2008-8} {\bibfield
  {journal} {\bibinfo  {journal} {Living Rev. Rel.}\ }\textbf {\bibinfo
  {volume} {11}},\ \bibinfo {pages} {8} (\bibinfo {year} {2008})},\ \Eprint
  {https://arxiv.org/abs/0811.0762} {arXiv:0811.0762 [astro-ph]} \BibitemShut
  {NoStop}%
\bibitem [{\citenamefont {Fragione}(2021)}]{Fragione:2021cvv}%
  \BibitemOpen
  \bibfield  {author} {\bibinfo {author} {\bibfnamefont {G.}~\bibnamefont
  {Fragione}},\ }\href {https://doi.org/10.3847/2041-8213/ac3bcd} {\bibfield
  {journal} {\bibinfo  {journal} {Astrophys. J. Lett.}\ }\textbf {\bibinfo
  {volume} {923}},\ \bibinfo {pages} {L2} (\bibinfo {year} {2021})},\ \Eprint
  {https://arxiv.org/abs/2110.09604} {arXiv:2110.09604 [astro-ph.HE]}
  \BibitemShut {NoStop}%
\bibitem [{\citenamefont {Qin}\ \emph {et~al.}(2019)\citenamefont {Qin},
  \citenamefont {Marchant}, \citenamefont {Fragos}, \citenamefont {Meynet},\
  and\ \citenamefont {Kalogera}}]{Qin:2018sxk}%
  \BibitemOpen
  \bibfield  {author} {\bibinfo {author} {\bibfnamefont {Y.}~\bibnamefont
  {Qin}}, \bibinfo {author} {\bibfnamefont {P.}~\bibnamefont {Marchant}},
  \bibinfo {author} {\bibfnamefont {T.}~\bibnamefont {Fragos}}, \bibinfo
  {author} {\bibfnamefont {G.}~\bibnamefont {Meynet}},\ and\ \bibinfo {author}
  {\bibfnamefont {V.}~\bibnamefont {Kalogera}},\ }\href
  {https://doi.org/10.3847/2041-8213/aaf97b} {\bibfield  {journal} {\bibinfo
  {journal} {Astrophys. J. Lett.}\ }\textbf {\bibinfo {volume} {870}},\
  \bibinfo {pages} {L18} (\bibinfo {year} {2019})},\ \Eprint
  {https://arxiv.org/abs/1810.13016} {arXiv:1810.13016 [astro-ph.SR]}
  \BibitemShut {NoStop}%
\bibitem [{\citenamefont {McClintock}\ \emph {et~al.}(2011)\citenamefont
  {McClintock}, \citenamefont {Narayan}, \citenamefont {Davis}, \citenamefont
  {Gou}, \citenamefont {Kulkarni}, \citenamefont {Orosz}, \citenamefont
  {Penna}, \citenamefont {Remillard},\ and\ \citenamefont
  {Steiner}}]{McClintock:2011zq}%
  \BibitemOpen
  \bibfield  {author} {\bibinfo {author} {\bibfnamefont {J.~E.}\ \bibnamefont
  {McClintock}}, \bibinfo {author} {\bibfnamefont {R.}~\bibnamefont {Narayan}},
  \bibinfo {author} {\bibfnamefont {S.~W.}\ \bibnamefont {Davis}}, \bibinfo
  {author} {\bibfnamefont {L.}~\bibnamefont {Gou}}, \bibinfo {author}
  {\bibfnamefont {A.}~\bibnamefont {Kulkarni}}, \bibinfo {author}
  {\bibfnamefont {J.~A.}\ \bibnamefont {Orosz}}, \bibinfo {author}
  {\bibfnamefont {R.~F.}\ \bibnamefont {Penna}}, \bibinfo {author}
  {\bibfnamefont {R.~A.}\ \bibnamefont {Remillard}},\ and\ \bibinfo {author}
  {\bibfnamefont {J.~F.}\ \bibnamefont {Steiner}},\ }\href
  {https://doi.org/10.1088/0264-9381/28/11/114009} {\bibfield  {journal}
  {\bibinfo  {journal} {Class. Quant. Grav.}\ }\textbf {\bibinfo {volume}
  {28}},\ \bibinfo {pages} {114009} (\bibinfo {year} {2011})},\ \Eprint
  {https://arxiv.org/abs/1101.0811} {arXiv:1101.0811 [astro-ph.HE]}
  \BibitemShut {NoStop}%
\bibitem [{\citenamefont {Foucart}(2012)}]{Foucart:2012nc}%
  \BibitemOpen
  \bibfield  {author} {\bibinfo {author} {\bibfnamefont {F.}~\bibnamefont
  {Foucart}},\ }\href {https://doi.org/10.1103/PhysRevD.86.124007} {\bibfield
  {journal} {\bibinfo  {journal} {Phys. Rev. D}\ }\textbf {\bibinfo {volume}
  {86}},\ \bibinfo {pages} {124007} (\bibinfo {year} {2012})},\ \Eprint
  {https://arxiv.org/abs/1207.6304} {arXiv:1207.6304 [astro-ph.HE]}
  \BibitemShut {NoStop}%
\bibitem [{\citenamefont {Foucart}\ \emph {et~al.}(2013)\citenamefont
  {Foucart}, \citenamefont {Deaton}, \citenamefont {Duez}, \citenamefont
  {Kidder}, \citenamefont {MacDonald}, \citenamefont {Ott}, \citenamefont
  {Pfeiffer}, \citenamefont {Scheel}, \citenamefont {Szilagyi},\ and\
  \citenamefont {Teukolsky}}]{Foucart:2012vn}%
  \BibitemOpen
  \bibfield  {author} {\bibinfo {author} {\bibfnamefont {F.}~\bibnamefont
  {Foucart}}, \bibinfo {author} {\bibfnamefont {M.~B.}\ \bibnamefont {Deaton}},
  \bibinfo {author} {\bibfnamefont {M.~D.}\ \bibnamefont {Duez}}, \bibinfo
  {author} {\bibfnamefont {L.~E.}\ \bibnamefont {Kidder}}, \bibinfo {author}
  {\bibfnamefont {I.}~\bibnamefont {MacDonald}}, \bibinfo {author}
  {\bibfnamefont {C.~D.}\ \bibnamefont {Ott}}, \bibinfo {author} {\bibfnamefont
  {H.~P.}\ \bibnamefont {Pfeiffer}}, \bibinfo {author} {\bibfnamefont {M.~A.}\
  \bibnamefont {Scheel}}, \bibinfo {author} {\bibfnamefont {B.}~\bibnamefont
  {Szilagyi}},\ and\ \bibinfo {author} {\bibfnamefont {S.~A.}\ \bibnamefont
  {Teukolsky}},\ }\href {https://doi.org/10.1103/PhysRevD.87.084006} {\bibfield
   {journal} {\bibinfo  {journal} {Phys. Rev. D}\ }\textbf {\bibinfo {volume}
  {87}},\ \bibinfo {pages} {084006} (\bibinfo {year} {2013})},\ \Eprint
  {https://arxiv.org/abs/1212.4810} {arXiv:1212.4810 [gr-qc]} \BibitemShut
  {NoStop}%
\bibitem [{\citenamefont {Ashton}\ \emph {et~al.}(2019)\citenamefont {Ashton}
  \emph {et~al.}}]{Ashton:2018jfp}%
  \BibitemOpen
  \bibfield  {author} {\bibinfo {author} {\bibfnamefont {G.}~\bibnamefont
  {Ashton}} \emph {et~al.},\ }\href {https://doi.org/10.3847/1538-4365/ab06fc}
  {\bibfield  {journal} {\bibinfo  {journal} {Astrophys. J. Suppl.}\ }\textbf
  {\bibinfo {volume} {241}},\ \bibinfo {pages} {27} (\bibinfo {year} {2019})},\
  \Eprint {https://arxiv.org/abs/1811.02042} {arXiv:1811.02042 [astro-ph.IM]}
  \BibitemShut {NoStop}%
\bibitem [{\citenamefont {{Cutler}}\ and\ \citenamefont
  {{Flanagan}}(1994)}]{1994PhRvD..49.2658C}%
  \BibitemOpen
  \bibfield  {author} {\bibinfo {author} {\bibfnamefont {C.}~\bibnamefont
  {{Cutler}}}\ and\ \bibinfo {author} {\bibfnamefont {{\'E}.~E.}\ \bibnamefont
  {{Flanagan}}},\ }\href {https://doi.org/10.1103/PhysRevD.49.2658} {\bibfield
  {journal} {\bibinfo  {journal} {\prd}\ }\textbf {\bibinfo {volume} {49}},\
  \bibinfo {pages} {2658} (\bibinfo {year} {1994})},\ \Eprint
  {https://arxiv.org/abs/gr-qc/9402014} {arXiv:gr-qc/9402014 [gr-qc]}
  \BibitemShut {NoStop}%
\bibitem [{\citenamefont {{Chassande-Mottin}}\ \emph
  {et~al.}(2019)\citenamefont {{Chassande-Mottin}}, \citenamefont {{Leyde}},
  \citenamefont {{Mastrogiovanni}},\ and\ \citenamefont
  {{Steer}}}]{2019PhRvD.100h3514C}%
  \BibitemOpen
  \bibfield  {author} {\bibinfo {author} {\bibfnamefont {E.}~\bibnamefont
  {{Chassande-Mottin}}}, \bibinfo {author} {\bibfnamefont {K.}~\bibnamefont
  {{Leyde}}}, \bibinfo {author} {\bibfnamefont {S.}~\bibnamefont
  {{Mastrogiovanni}}},\ and\ \bibinfo {author} {\bibfnamefont {D.~A.}\
  \bibnamefont {{Steer}}},\ }\href
  {https://doi.org/10.1103/PhysRevD.100.083514} {\bibfield  {journal} {\bibinfo
   {journal} {\prd}\ }\textbf {\bibinfo {volume} {100}},\ \bibinfo {eid}
  {083514} (\bibinfo {year} {2019})},\ \Eprint
  {https://arxiv.org/abs/1906.02670} {arXiv:1906.02670 [astro-ph.CO]}
  \BibitemShut {NoStop}%
\bibitem [{\citenamefont {{Wang}}\ \emph {et~al.}(2022)\citenamefont {{Wang}},
  \citenamefont {{Tang}}, \citenamefont {{Li}}, \citenamefont {{Jin}},\ and\
  \citenamefont {{Fan}}}]{2022PhRvD.106b3011W}%
  \BibitemOpen
  \bibfield  {author} {\bibinfo {author} {\bibfnamefont {Y.-Y.}\ \bibnamefont
  {{Wang}}}, \bibinfo {author} {\bibfnamefont {S.-P.}\ \bibnamefont {{Tang}}},
  \bibinfo {author} {\bibfnamefont {X.-Y.}\ \bibnamefont {{Li}}}, \bibinfo
  {author} {\bibfnamefont {Z.-P.}\ \bibnamefont {{Jin}}},\ and\ \bibinfo
  {author} {\bibfnamefont {Y.-Z.}\ \bibnamefont {{Fan}}},\ }\href
  {https://doi.org/10.1103/PhysRevD.106.023011} {\bibfield  {journal} {\bibinfo
   {journal} {\prd}\ }\textbf {\bibinfo {volume} {106}},\ \bibinfo {eid}
  {023011} (\bibinfo {year} {2022})},\ \Eprint
  {https://arxiv.org/abs/2111.02027} {arXiv:2111.02027 [astro-ph.HE]}
  \BibitemShut {NoStop}%
\bibitem [{\citenamefont {Nesseris}\ \emph {et~al.}(2013)\citenamefont
  {Nesseris}, \citenamefont {Basilakos}, \citenamefont {Saridakis},\ and\
  \citenamefont {Perivolaropoulos}}]{Nesseris:2013jea}%
  \BibitemOpen
  \bibfield  {author} {\bibinfo {author} {\bibfnamefont {S.}~\bibnamefont
  {Nesseris}}, \bibinfo {author} {\bibfnamefont {S.}~\bibnamefont {Basilakos}},
  \bibinfo {author} {\bibfnamefont {E.~N.}\ \bibnamefont {Saridakis}},\ and\
  \bibinfo {author} {\bibfnamefont {L.}~\bibnamefont {Perivolaropoulos}},\
  }\href {https://doi.org/10.1103/PhysRevD.88.103010} {\bibfield  {journal}
  {\bibinfo  {journal} {Phys. Rev. D}\ }\textbf {\bibinfo {volume} {88}},\
  \bibinfo {pages} {103010} (\bibinfo {year} {2013})},\ \Eprint
  {https://arxiv.org/abs/1308.6142} {arXiv:1308.6142 [astro-ph.CO]}
  \BibitemShut {NoStop}%
\bibitem [{\citenamefont {Perivolaropoulos}\ and\ \citenamefont
  {Skara}(2022)}]{Perivolaropoulos:2022khd}%
  \BibitemOpen
  \bibfield  {author} {\bibinfo {author} {\bibfnamefont {L.}~\bibnamefont
  {Perivolaropoulos}}\ and\ \bibinfo {author} {\bibfnamefont {F.}~\bibnamefont
  {Skara}},\ }\href {https://doi.org/10.3390/universe8100502} {\bibfield
  {journal} {\bibinfo  {journal} {Universe}\ }\textbf {\bibinfo {volume} {8}},\
  \bibinfo {pages} {502} (\bibinfo {year} {2022})},\ \Eprint
  {https://arxiv.org/abs/2208.11169} {arXiv:2208.11169 [astro-ph.CO]}
  \BibitemShut {NoStop}%
\bibitem [{\citenamefont {Aubourg}\ \emph {et~al.}(2015)\citenamefont {Aubourg}
  \emph {et~al.}}]{Aubourg:2014yra}%
  \BibitemOpen
  \bibfield  {author} {\bibinfo {author} {\bibfnamefont {E.}~\bibnamefont
  {Aubourg}} \emph {et~al.},\ }\href
  {https://doi.org/10.1103/PhysRevD.92.123516} {\bibfield  {journal} {\bibinfo
  {journal} {Phys. Rev. D}\ }\textbf {\bibinfo {volume} {92}},\ \bibinfo
  {pages} {123516} (\bibinfo {year} {2015})},\ \Eprint
  {https://arxiv.org/abs/1411.1074} {arXiv:1411.1074 [astro-ph.CO]}
  \BibitemShut {NoStop}%
\bibitem [{\citenamefont {Foreman-Mackey}\ \emph {et~al.}(2013)\citenamefont
  {Foreman-Mackey}, \citenamefont {Hogg}, \citenamefont {Lang},\ and\
  \citenamefont {Goodman}}]{foreman2013emcee}%
  \BibitemOpen
  \bibfield  {author} {\bibinfo {author} {\bibfnamefont {D.}~\bibnamefont
  {Foreman-Mackey}}, \bibinfo {author} {\bibfnamefont {D.~W.}\ \bibnamefont
  {Hogg}}, \bibinfo {author} {\bibfnamefont {D.}~\bibnamefont {Lang}},\ and\
  \bibinfo {author} {\bibfnamefont {J.}~\bibnamefont {Goodman}},\ }\href@noop
  {} {\bibfield  {journal} {\bibinfo  {journal} {Publications of the
  Astronomical Society of the Pacific}\ }\textbf {\bibinfo {volume} {125}},\
  \bibinfo {pages} {306} (\bibinfo {year} {2013})}\BibitemShut {NoStop}%
\bibitem [{\citenamefont {{Lewis}}(2019)}]{2019arXiv191013970L}%
  \BibitemOpen
  \bibfield  {author} {\bibinfo {author} {\bibfnamefont {A.}~\bibnamefont
  {{Lewis}}},\ }\href {https://doi.org/10.48550/arXiv.1910.13970} {\bibfield
  {journal} {\bibinfo  {journal} {arXiv e-prints}\ ,\ \bibinfo {eid}
  {arXiv:1910.13970}} (\bibinfo {year} {2019})},\ \Eprint
  {https://arxiv.org/abs/1910.13970} {arXiv:1910.13970 [astro-ph.IM]}
  \BibitemShut {NoStop}%
\end{thebibliography}%
\end{document}